\documentclass[structabstract]{aa}
\usepackage{natbib}
\usepackage{graphicx,xcolor}
\usepackage[normalem]{ulem}
\usepackage{natbib}
\usepackage{textcomp} %\textmu
\usepackage{txfonts,mathrsfs}

\def\elogP{$e - \log P$}
\def\chem#1{$^{#1}$}
\def\msun{\,$M_\odot$}
\def\Msun{M_\odot}
\def\Myr{M_\odot\,\mathrm{yr}^{-1}}
\def\rsun{\,$R_\odot$}
\def\Rsun{\,R_\odot}

\def\RLOF{\mathrm{RLOF}}
\def\WRLOF{\mathrm{WRLOF}}
\def\Lone{{L_1}}

\def\Ltwo{{L_2}}
\def\acc{\mathrm{acc}}
\def\dust{\mathrm{dust}}
\def\iso{\mathrm{iso}}
\def\orb{\mathrm{orb}}
\def\mmax{\mathrm{max}}
\def\crit{\mathrm{crit}}
\def\wind{\mathrm{wind}}

\def\circ{\mathrm{circ}}
\def\tides{\mathrm{tides}}
\def\tide{\mathrm{tide}}

\def\CB{\mathrm{CB}}
\def\BH{\mathrm{BHL}}
\def\inn{\mathrm{in}}

\def\dust{\mathrm{dust}}

\def\res{\mathrm{res}}
\def\visc{\mathrm{visc}}

\def\gcm{g\,cm$^{-2}$}
\def\kms{km\,s$^{-1}$}
\def\RdRL{R_{\mathrm{dust}}/R_{\mathrm{L}_1}}

\begin{document}
   \title{Formation of Ba stars : impact of wind Roche lobe overflow and circumbinary disk in
     shaping the orbital parameters}

   \author{P. Krynski\inst{1}
          \and
          L. Siess\inst{1}
          \and
          A. Jorissen\inst{1}
          \and
          P. J. Davis\inst{2}
          }

\institute{Institut d'Astronomie et d'Astrophysique and
BLU-ULB (Brussels Laboratory of the Universe), Universit\'{e}
Libre de Bruxelles (ULB), Avenue F.D. Roosevelt 50, CP226, Brussels, Belgium
\and
Laboratoire de physique corpusculaire - Caen, France}

\titlerunning{Formation of Ba stars}

%   \date{Received September 15, 1996; accepted March 16, 1997}

  \abstract
  {After more than three decades of investigation, the distribution of Ba stars in the \elogP{}  diagram still defies our understanding. Recent smooth particle hydrodynamic simulations involving an asymptotic giant branch (AGB) primary have shown that a circumbinary disk (CB) can form around the binary and that the presence of dust in the wind of evolved low- and intermediate-mass stars can significantly affect the systemic angular momentum loss and mass accretion onto the companion through the wind Roche lobe overflow (WRLOF) phase.}
  {The aim of this paper is to provide a more realistic modeling of the progenitors of Ba stars and investigate the role of various processes (CB disk, WRLOF, tidally enhanced wind mass loss, and non-conservative RLOF) on the orbital evolution.}
  {We used the binary evolution code {\sf{BINSTAR}}, which includes the latest prescriptions for mass transfer in the WRLOF regime. In our approach, we considered that a CB disk forms when WRLOF is activated. The coupling between the CB disk and the binary follows the standard resonant interaction theory. We constructed grids of 2.0 + 1.0~\msun{} and 1.2 + 0.8~\msun{} binaries for initial orbital parameters that result in WRLOF, and evolved these systems until the end of the primary's AGB phase. A tentative approach was made to estimate the effect of the CB disk during the post-AGB phase.}
  {WRLOF resulted in a significant shrinkage of the orbital separation during the AGB phase, leading to binaries with initial periods on the order of $\lesssim 12000$~d undergoing Roche lobe overflow (RLOF). The combination of WRLOF, eccentricity pumping from the CB disk, and/or tidally enhanced wind mass loss can lead to RLOF on eccentric orbits down to periods of $P_\orb \sim 3000$~d. Non-conservative RLOF enabled a reduction of the period before circularization down to $\sim 2000$~d, provided at least 50 percent of the transferred mass left the system.}
  {Our models still cannot account for the eccentricity distribution of Ba stars with periods shorter than $P_\orb \lesssim 2000$~d, where a common envelope evolution appears unavoidable.}


   \mail{pawel.krynski@ulb.be}

   \maketitle

\section{Introduction}
Approximately 50 percent of low- and intermediate-mass stars ($1.5~\Msun \lesssim M \lesssim 5~\Msun$) in the Galaxy are part of a binary system \citep{Duchene2013}, where interactions between the two components can impact their evolution and orbital parameters. An example of such systems is barium (Ba) stars, which are main-sequence or giant stars with surface enrichment in heavy elements produced during the so-called s-process \citep{1998A&A...332..877J,swaelmen2017}. Nearly all Ba stars are found in binary systems with a white dwarf (WD) companion, leading to the widely accepted paradigm that the observed pollution in heavy elements occurred when the WD was on the asymptotic giant branch (AGB) and transferred the products of its nucleosynthesis to its binary companion \citep[for example,][]{McClure1980,McClure1984,jorissen2019}. As the AGB companion evolves into a faint WD, only the Ba star remains detectable.
The mass transfer responsible for the chemical pollution is also likely to affect the orbital properties of these systems. Observations of Ba stars show that systems with orbital periods less than 2000–3000 days are able to maintain a high eccentricity, at odds with current predictions from tidal models \citep{Zahn1977}, which predict efficient circularization.
These systems inherited their orbital properties from past binary interactions that most likely occurred during the AGB phase, when the primary reached its largest extent and was ejecting its envelope via dust-driven winds.

Hydrodynamical simulations of binary systems involving a primary AGB star \citep[for example,][]{TheunsI,MastrodemosI,Kim2012A,Maes2021} have shown that if the AGB wind velocity is slower than the orbital speed, the gravitational pull of the companion star alters the shape of the outflow. The wind outflow shifts from an isotropic morphology to one that is directed toward the companion star, and for this resemblance to RLOF it earned the name wind Roche lobe overflow \citep[WRLOF;][]{MohamedPodsiadlowski2007}. The modified outflow pattern results in an increased accretion efficiency onto the companion star. Furthermore, winds may escape through the second Lagrangian point ($\Ltwo$), carrying an increased amount of angular momentum, leading to a shortening of the orbital period, contrary to the case of isotropic, fast winds \citep{saladino2019,chen2018}.
\cite{Abate2018} argued that increased angular momentum extraction by winds could be an important factor explaining the orbital properties of CEMP-s stars, the low-metallicity counterparts of Ba stars.

Winds escaping through $\Ltwo$ also give rise to the formation of a spiral shock behind the companion star, potentially evolving into a circumbinary disk (CB disk) \citep{MohamedPodsiadlowski2007,chen2018}.
Moreover, the spectral energy distributions of all post-AGB binaries indicate the presence of a Keplerian CB disk \citep{bujarrabal2013,oomen2018}.
The interaction between the binary system and the CB disk is commonly modeled using linear perturbation theory involving Lindblad resonances, which are known to induce eccentricity in the orbits \citep{1979ApJ...233..857G,1994ApJ...421..651A}.
However, many studies have struggled to explain the eccentricities of post-AGB binaries with periods below 1000~days without assuming disk masses that exceed observed values \citep{Rafikov2016,oomen2020}.
Although no better model for long-term binary–CB disk interaction currently exists, several recent smoothed-particle hydrodynamics (SPH) simulations suggest that the process may be more complex, involving accretion streams from the CB disk that enter the binary cavity, leading to mass deposition on both binary components \citep{munoz2019,tiede2020,orazio2021,zrake2021}.

When initiated on an eccentric orbit, RLOF mass transfer is phase dependent, increasing near periastron and decreasing near apastron. This phase dependence can generate eccentricity that effectively counteracts the circularization induced by tidal forces. \cite{vos2015} suggested that this principle could explain the properties of long-period sdB binaries.
A similar mechanism, known as tidally enhanced wind mass loss, proposed by \cite{ToutEggleton1988}, introduces a phase dependency in the wind mass loss, resulting in an increased ejection rate at periastron. Akin to the phase-dependent RLOF, this modified wind mass loss can generate eccentricity in binary systems. Furthermore, the enhanced mass loss accelerates the evolutionary pace of the AGB star, reducing the time available for tidal forces to circularize the orbit. Consequently, the star's evolution may terminate earlier with a smaller radius, enabling it to reach shorter orbital periods before initiating RLOF. This mechanism was invoked by \cite{2014A&A...565A..57S} to explain the properties of the He white dwarf system IP Eri. Additionally, \cite{Bonavic2008} utilized this mechanism, coupled with phase-dependent RLOF, to elucidate the origin of eccentric post-AGB binaries.

This study provides one of the first investigations into the long-term impact of WRLOF, CB disk, and tidally enhanced wind mass loss on the orbital evolution of potential progenitor systems of Ba stars. To achieve this, we use the binary evolution code {\sf BINSTAR} \citep{2013A&A...550A.100S}, which incorporates these key mechanisms.
The paper is organized as follows: In Sect.~2, we present our modeling of WRLOF, phase-dependent RLOF, and our model of the CB disk and its interaction with the binary. In Sect.~3, we estimate the range of initial orbital parameters for the systems to experience WRLOF. In Sect.~4, we present our results, beginning with an overview of the observational data sample, followed by an analysis of a representative system. We then present the evolutionary tracks in the \elogP{} diagram for a particular set of parameters, followed by an exploration of the effects of varying these parameters. We also analyze the effect of tidally enhanced wind mass loss and of changing the initial mass of the primary. Finally, we estimate the effects of the CB disk after the end of the AGB phase until its dissipation, and the effect of non-conservative RLOF. In Sect.~5, we discuss our findings, and in Sect.~6, we provide a conclusion.
\section{The binary paradigm}

\subsection{The Wind Roche Lobe Overflow (WRLOF)}
\label{disc_formation}

Hydrodynamical simulations of the evolution of an AGB primary with a lower-mass companion \citep[for example,][]{MohamedPodsiadlowski2007,deValBorro2009,chen2017,saladino2019} indicate that if dust forms near or beyond the AGB's Roche lobe (i.e., $R_\dust/R_{\Lone} \gtrsim 1$), then the radiatively driven wind is not accelerated to the AGB's escape velocity. Consequently, the wind is not ejected isotropically, but channels through the inner Lagrangian ($\Lone$) point and focuses onto the companion star. The flow geometry resembles the classical RLOF scenario and has since been coined wind Roche lobe overflow.
A fraction of the wind is accreted onto the companion, leading to a significant increase in the companion's mass.
The remaining material escapes the system through the $\Ltwo$ point behind the AGB's companion, carrying more angular momentum compared to the isotropic wind mass loss case, producing a shrinkage of the orbit \citep{Jahanara2015,chen2018,saladino2019}.
\citet{2012BaltA..21...88M} also found that the gas escaping through $\Ltwo$ leads to the formation of a single spiral arm around the binary. The spiral is compressed, and the enhanced density can trigger dust formation. Whether this structure subsequently evolves into a stable CB disk has been claimed \citep{chen2017,chen2020}.

In this study, our paradigm is that WRLOF leads to the formation of a CB disk as a result of material channeling through the $\Ltwo$ point. In our models, we assume that WRLOF and disk formation occur once $R_\dust/R_{\Lone} > 1$.
We estimate the dust formation radius $R_\dust$ using the formulation of \citet{2007ASPC..378..145H}: \begin{equation} R_\dust = \frac{1}{2} R_{1} \left(\frac{T_{\mathrm{eff,1}}}{T_\dust}\right)^{2.5}, \label{eq:Rdust} \end{equation} where $T_{\mathrm{eff,1}}$ and $R_{1}$ are the AGB's effective temperature and radius, and $T_\dust$ the dust condensation temperature. For carbon-rich ($\mathrm{C}/\mathrm{O} > 1$) stars, $T_\dust \approx 1500$~K, and for oxygen-rich stars (C/O $< 1$), $T_\dust \approx 1000$~K.

\subsection{Evolution of the orbital parameters}

The change in the orbital separation is determined from the rates of change of the donor and gainer masses, $\dot{M}_{1}$ and $\dot{M}_{2}$ respectively, the orbital angular momentum loss rate, $\dot{J}_\orb$, and the rate of change of the eccentricity $\dot{e}$, i.e.
\begin{equation}
\frac{\dot{a}}{a}=2\frac{\dot{J}_\orb}{J_\orb}-2\frac{\dot{M}_{1}}{M_{1}}
-2\frac{\dot{M}_{2}}{M_{2}}+\frac{\dot{M}_{1}+\dot{M}_{2}}{M_1+M_2}+\frac{2e\dot{e}}{1-e^{2}},
\label{eq:adot_a}
\end{equation}
where
\begin{equation}
\dot{e} = \dot{e}_{\tides} + \dot{e}_\res + \dot{e}_\mathrm{winds}+ \dot{e}_\mathrm{RLOF}
\label{eq:edot}
\end{equation}
is the sum of the contributions from tides $\dot{e}_{\tides}$, calculated using the prescriptions of \citet{Zahn1977,1989A&A...220..112Z}, the resonant interactions with the CB disk $\dot{e}_\res$ (Eq.~\ref{edot_2}) and the exchanges of mass between the stars by the winds $\dot{e}_\mathrm{winds}$ and RLOF $\dot{e}_\mathrm{RLOF}$ (Eq.~\ref{eq:edot_exRLOF}).

The total angular momentum of the binary, $J_{\Sigma}$, is the sum of the orbital angular momentum, $J_\orb$ and the spin angular momenta of each star, $J_{1,2}$ 
\begin{equation}
  J_{\Sigma}=J_\orb+J_{1}+J_{2} 
\label{J_tot}
\end{equation}
Taking the time derivative of Eq.~\ref{J_tot} and solving for $\dot{J}_\orb$ gives
\begin{equation}
  \dot{J}_\orb=-\dot{J}_{1}-\dot{J}_{2}+\dot{J}_{\Sigma}
\label{Jorb_dot}
\end{equation}
The torques on the stellar spins $\dot{J}_{1,2}$ come from tidal interaction and mass transfer and are given in \cite{2013A&A...550A.100S}.

\subsection{Wind mass transfer}

In the WRLOF regime, the wind accretion efficiency onto the companion, $\beta_\wind = |\dot{M}_{\acc,2}/\dot{M}_1|$, and the angular momentum carried away from the system by the wind, $\dot{J}_{\wind}$, are increased compared to the standard (non-dusty) prescription. To parameterize $\dot{J}_\Sigma$, we use the prescription derived by \citet{saladino2019} based on SPH simulations.
In their formulation, the systemic angular momentum loss is given by
\begin{equation}
    \dot{J}_{\Sigma} = \eta \frac{J_{\orb}}{\mu} \dot{M}
    \label{eq:Jorbdot_WRLOF}
\end{equation}
where $\mu= M_1M_2/(M_1+M_2)$ is the reduced mass and $\eta$ a function of the ratio of the wind to orbital velocities ($v_{\wind}/v_{\orb}$) and of the mass ratio $q=M_1/M_2$ given by
\begin{equation}
\eta \left(\frac{v_{\wind}}{v_{\orb}},q\right) = \mathrm{min}\left(
\frac{1}{c_1+(c_2 v_{\wind}/v_{\orb})^3}+\eta_\iso,0.6 \right),
\label{eq:eta}
\end{equation}
where $c_1 = \mathrm{max}(q,0.6 q^{1.7})$,  $c_2 = 1.5+0.3q$ and $\eta_{\iso} = 1/(1+q)^2$ is the fraction of the specific orbital angular momentum lost by the wind in the isotropic (fast) wind case.

We can assess the effect of fast winds on the orbital period by substituting $\dot{J}\orb$ in Eq.~\ref{eq:adot_a} with $\dot{J}_\Sigma$ in Eq.~\ref{eq:Jorbdot_WRLOF}, and neglecting accretion onto the companion, $\dot{M}_2 = 0$.
For a circular orbit, we find
\begin{equation}
\frac{\dot{a}}{a} = \frac{2\: \dot{M}}{q \: M} \left( \eta \:  (1+q)^2 - (1+\frac{q}{2}) \right).
\label{eq:adot_a_sys}
\end{equation}
In the limit $v_\wind \gg v_\orb$, $\eta$ approaches the fast (isotropic) wind case  $\eta_\iso$, resulting in
\begin{equation}
\frac{\dot{a}}{a} = -\frac{\dot{M_1}}{M},
\end{equation}
showing that the orbit widens as the primary loses mass, independently of the mass ratio.
Following \citet{2014A&A...565A..57S}, this equation can be integrated, and using Kepler's third law, the evolution of the orbital period can be expressed in terms of the mass ratio
\begin{equation}
P = P_0 \left(\frac{1+q_0}{1+q}\right)^2,
\label{eq:PeriodEvolution}
\end{equation}
where $P$, $P_0$, $q$ and $q_0$ are the current and initial period and mass ratio, respectively.
Equation \ref{eq:adot_a_sys} also tells us that if $\eta > \eta_{\crit} = (1+\frac{q}{2})\,\eta_{\iso}$, systemic wind mass loss produces a shrinkage of the orbit.

\begin{figure}
  \begin{center}
    \includegraphics[width = \columnwidth]{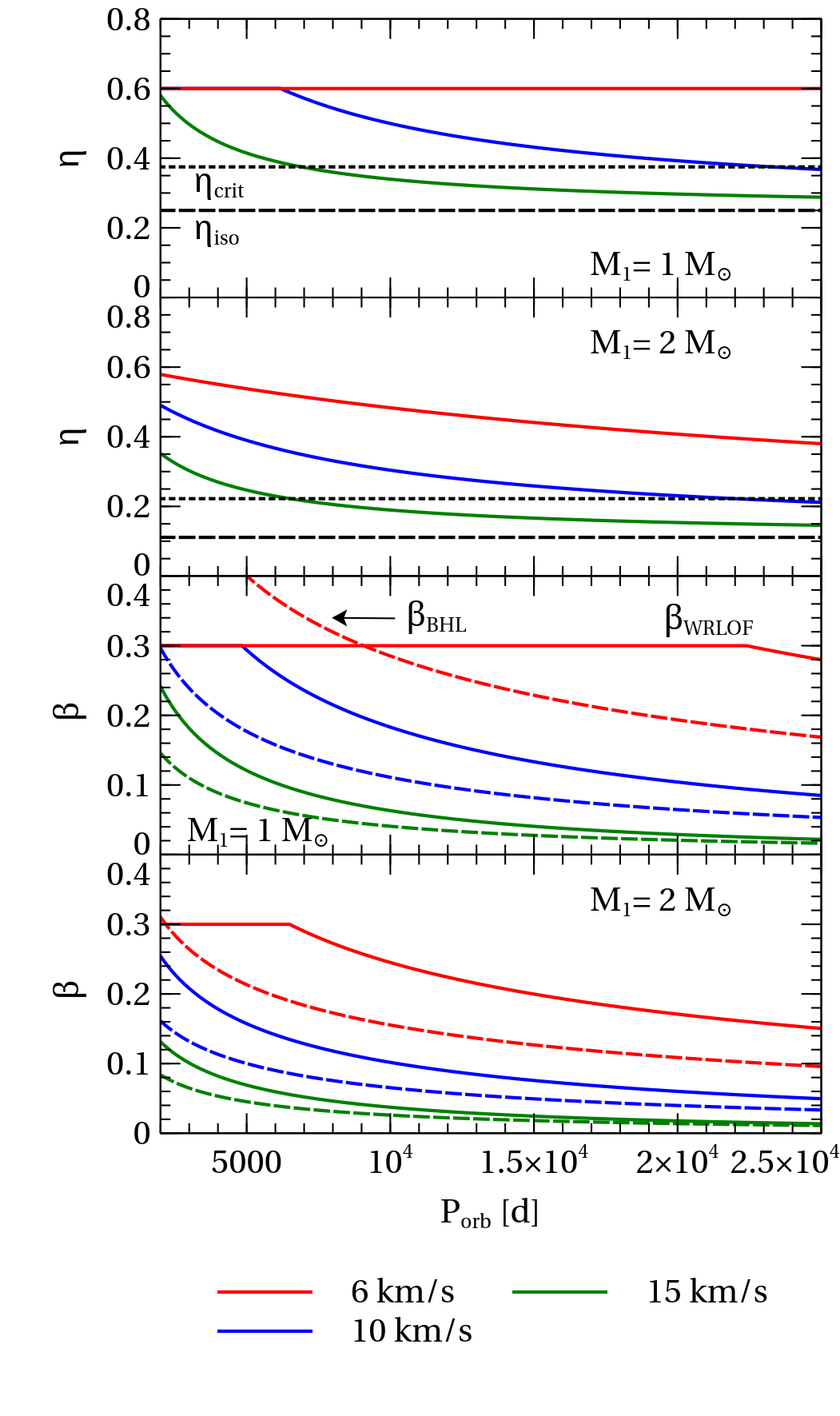}
    \caption{{\it Top two panels}: Fraction of the specific angular momentum $\eta$ carried away by winds during WRLOF as a function of orbital period for a $1+1$\msun{} (top panel) and $2+1$\msun{} (top-middle panel) binary system, and for different wind velocities ($v_\wind$, as labeled by the different colors). The horizontal long-dashed line represents the value of $\eta$ in the case of isotropic wind emission (Jeans mode) and the short-dashed line the value of $\eta_\crit$ above which the orbit shrinks as a result of wind mass loss. {\it Bottom two panels}: Accretion efficiency onto the companion, $\beta= |\dot{M}_{\acc,2}/\dot{M}_1|$ for the same two systems as in the upper panels, either through wind ($\beta_{\mathrm{BHL}}$, dashed lines) or through WRLOF ($\beta_{\mathrm{WRLOF}}$, solid lines).
    }
    \label{fig:eta_parameter}
  \end{center}
\end{figure}

The top two panels of Fig.~\ref{fig:eta_parameter} show the dependence of $\eta$ on the orbital period for two systems characterized by ($M_1,\ M_2$) = (2,1) and (1,1), respectively (with masses expressed in \msun). The wind velocity is a free parameter, and for AGB stars, we use representative values of 6, 10 and 15~\kms. For both values of $q$ and for $v_\wind = 6$ and 10~\kms, systems with $P_\orb < 20000$~d have $\eta >\eta_\crit$, meaning that systemic wind mass loss will tighten the orbit. For $v_\wind = 15$~\kms, $\eta < \eta_{\crit}$ only for systems with periods $\gtrsim 6000$~d.
In wider binaries, $v_\wind \gg v_\orb$ so the fast wind regime prevails and mass loss pushes the stars apart.

The wind accretion efficiency onto the companion is generally estimated from the Bondi-Hoyle-Littleton prescription \citep[e.g.][]{1988A&A...205..155B}
\begin{equation}
  \beta_\BH=\frac{\alpha_{\BH}}{(1-e^{2})^{\frac{1}{2}}(1+q)^2}
  \left(\frac{v_\orb}{v_\wind}\right)^{4}\left[1+
    \left(\frac{v_\orb}{v_\wind}\right)^{2}\right]^{\,-\frac{3}{2}},
\label{BHL}
\end{equation}
where $\alpha_\BH=0.75$ is a parameter, $v_\orb=(GM/a)^{\frac{1}{2}}$ the average orbital velocity, and $v_\wind$ is the wind velocity given by Eq.~3 of \citet{1993ApJ...413..641V}. This prescription should be restricted to fast winds, where $v_\wind\gg v_\orb$.
As shown in hydrodynamical simulations, when $v_\wind \lesssim v_\orb$, the wind is focused towards the orbital plane, leading to an increased accretion efficiency onto the companion. Based on their SPH simulations, \citet{saladino2019} derived a prescription for the accretion efficiency $\beta_\WRLOF$ onto the companion that is given by
\begin{equation}
\beta_\WRLOF = \mathrm{min}(\alpha \beta_{\BH}, \beta_\mmax),
\label{eq:beta}
\end{equation}
where $\beta_\BH$ is computed using Eq.~\ref{BHL} with $\alpha_\BH = 1$, and
\begin{equation}
\alpha = 0.75 + \frac{1}{k_1+(k_2 v_\wind/v_\orb)^5},
\end{equation}
where $k_1 = 1.7 +0.3\:q$, $k_2 = 0.5+0.2\:q$, and $\beta_\mmax = \mathrm{min}(0.3,1.4\: q^{-2})$, the maximal accretion efficiency found in their simulations.
The bottom two panels in Fig.~\ref{fig:eta_parameter} show $\beta_{\mathrm{WRLOF}}$ and $\beta_{\mathrm{BHL}}$ as a function of the orbital period for two different mass ratios and different wind velocities. As expected, the accretion efficiency decreases for higher values of $v_\wind$ and in systems with longer orbital periods, where $v_\orb$ is slower.

Variations of the mass loss rate along an elliptical orbit induce a change in eccentricity that can be expressed as  \citep{2011epbm.book.....E}
\begin{eqnarray}
\dot{e}_\wind &=& \Bigg[ \left(1-\beta_{\wind}\right) \frac{|\dot{M}_{\wind}|}{M} + 2 \: \beta_{\wind} \: |\dot{M}_{\wind}|\left(\frac{1}{M_1}-\frac{1}{M_2}\right)\Bigg] \nonumber \\
  & &   \times \,(e+\cos\nu),
\label{eq:edot_exW}
\end{eqnarray}
where $\nu$ is the true anomaly and $\beta_{\wind}$ the efficiency of wind accretion onto the companion. When the evolutionary timestep, $\Delta t > P_\orb$, Eq.~\ref{eq:edot_exW} is averaged over the orbital period using the procedure outlined in \citet{2014A&A...565A..57S}. The net change in eccentricity is nonzero only if the bracketed term depends on the orbital phase. This happens for example for tidally enhanced winds (Sect.~\ref{impact_crap}) and, in this case, the systemic mass loss (first term in brackets) always leads to eccentricity pumping, while wind accretion onto the companion (second term) has a positive contribution only when $M_1 < M_2$ and negative in the opposite case. The total change in eccentricity due to wind mass loss should include the contributions from both stellar components.

\subsection{Change in $a$ and $e$ due to Roche lobe overflow}

As the primary ascends its AGB, it may eventually fill its Roche lobe and initiate RLOF. The expansion of the convective envelope of the AGB upon mass loss may lead to dynamically unstable mass transfer and common envelope evolution, unless the Roche lobe expands at a similar rate as the star's radius.

Assuming conservative RLOF ($\dot{M}_1  = \dot{M}_\RLOF= -\dot{M}_2$ and $\dot{J}_\orb =0$), Eq.~\ref{eq:adot_a} writes
\begin{equation}
    \frac{\dot{a}}{a} =   2\: \dot{M}_\RLOF \left(\frac{1}{M_1}-\frac{1}{M_2}\right) + \frac{2\: e \: \dot{e}}{1-e^2}.
\label{eq:adot_RLOF}
\end{equation}
If the orbit is eccentric, the change in $e$ due to mass loss is also given by Eq.~\ref{eq:edot_exW}, replacing $\dot{M}_\wind$ by $\dot{M}_\RLOF$ and $\beta_\wind$ by $\beta_{\mathrm{RLOF}} = -\dot{M}_1/\dot{M}_2$ ($=1$ in our conservative approach), to give (see also \citealt{Soker2000})
\begin{equation}
\dot{e}_\mathrm{RLOF} = \Bigg[
                2 \dot{M}_\RLOF \left(\frac{1}{M_{1}}-\frac{1}{M_{2}}\right) \Bigg]
                \times \,(e+\cos\nu).
  \label{eq:edot_exRLOF}
\end{equation}
It follows that if RLOF occurs when $M_1 >M_2$, both the separation and the eccentricity decrease (the second term in Eq.~\ref{eq:adot_RLOF} is generally negligible for this conclusion to hold).
However, when the mass ratio is reverted, RLOF widens the orbit and the eccentricity increases.

\subsection{Resonant interactions with the circumbinary disk}
\label{sec:resonances}

The resonant interactions between the CB disk and the binary have been studied by \cite{1979ApJ...233..857G} and \cite{1994ApJ...421..651A} using linear perturbation theory. They found that the binary can exert a positive or negative torque on the CB disk at resonant locations, which, in turn, affects the orbital parameters.
The potential of the binary $\Phi(r,\theta,t)$ (where $r$ is the distance from the CB disk center located at the binary's mass center, and $\theta$ is the azimuthal angle) is developed in series as
\begin{equation}
  \Phi(r,\theta,t)=\sum_{m\ell}\phi_{m\ell}(r)\exp[i(m\theta-\ell\Omega_{\mathrm{B}}t)],
\label{phi}
\end{equation}
where $l$ is the time-harmonic number, $m \geq 0$ is the azimuthal number, $\phi_{m\ell}(r)$ is the radially varying potential component for a given $(m,\ell)$ and $\Omega_{\mathrm{B}}=2\pi/P_\orb$ is the binary's mean orbital angular speed, with $P_\orb$ the orbital period.
Each harmonic $\phi_{m\ell}(r)$ is associated with an inner Lindblad resonance (ILR) that produces a positive torque on the binary and an outer Lindblad resonance (OLR) that exerts a negative torque. The resonances are located at radii
\begin{equation}
    r_{LR} = \left(\frac{m\pm1}{l}\right)^{\frac{2}{3}} a,
\end{equation}
where the positive and negative signs refer to the OLR and ILR, respectively, and $a$ is the binary semi-major axis.

For small eccentricities $(e \lesssim 0.2)$, SPH simulations \citep{1994ApJ...421..651A} show that the $(2,1)$ OLR is the dominant component for the angular momentum exchange between the CB disk and the binary.
From now on, we write results relevant only to this resonance. A more general form of the equations, valid for any other resonance is presented in Appendix A.
\cite{1979ApJ...233..857G} computed the torque $\dot{J}_\res$ exerted by the CB on the binary assuming that ($i$) the disk is thin ($H/r \lesssim 0.1$, where $H$ is the disk's pressure scale height), ($ii$) the disk is coplanar with the binary orbit, and ($iii$) the potential of the non-axisymmetric perturbations of the disk's structure is small compared to that of the binary.
The torque, evaluated at the $(2,1)$ outer resonance, and in the limit of small $e$ reads
\begin{equation}
  \dot{J}_\res= - \frac{49 \pi^2}{8} G\, \frac{\mu^2}{M} a\,  \Sigma(r_\res)\, e^2,
\label{Jdot_ml}
\end{equation}
where $\Sigma(r_\res)$ is the disk's surface density at the $(2,1)$ OLR located at $r_\res=3^{2/3}a \approx 2.08 \: a $.
The effect of the torque on the binary separation writes
\begin{equation}
 \left(\frac{\dot{a}}{a}\right)_\res=-\frac{49 \pi^2}{8} \Sigma(r_\res)\frac{\mu}{M} \sqrt{G a} \:  e^2 \: ,
\label{adot_a_CB}
\end{equation}
where $\mu=M_{1}M_{2}/(M_{1}+M_{2})$ is the reduced mass.
Finally, the associated rate of change of the eccentricity is
\begin{equation}
\dot{e}_\res= -\left(\frac{1-e^{2}}{e}\right)
\left[\frac{1}{2}-\frac{1}{(1-e^{2})^{\frac{1}{2}}}\right]\left(\frac{\dot{a}}{a}\right)_\res.
\label{edot_2}
\end{equation}
The $(2,1)$ OLR acts to shrink the orbit and increase the eccentricity of the binary. Substituting Eq.~\ref{adot_a_CB} into Eq.~\ref{edot_2}, it becomes apparent that the eccentricity pumping is proportional to $e$ and vanishes for circular orbits.

In more eccentric systems, higher order OLRs and ILRs are activated. The inner Lindblad resonances have the opposite effect on the orbit than the outer ones, and taking them into account, the eccentricity pumping is damped. \citet{2011MNRAS.415.3033R} found that for black hole binaries with $e \approx 0.6-0.8$ the eccentricity pumping roughly cancels and $e$ saturates. A similar behavior is also found in the 2D SPH simulations performed by \cite{orazio2021}, where they explain this saturation by the deformation of the disk when $e>0.4$ \citep[see also][]{Valli2024}. We account for the damping effects of the LRs at high eccentricities by multiplying Eqs.(\ref{Jdot_ml}) and (\ref{adot_a_CB}) by a Fermi-like function, $\mathcal{F}(e)$, given by
\begin{equation}
  \mathcal{F}(e)=\frac{1}{\exp\left(\frac{e-0.3}{0.04}\right)+1}.
\label{e_damp}
\end{equation}
This allows us to reproduce the saturation effect found by \citet{Valli2024}, i.e. $\dot{e}\rightarrow 0$ when $e > 0.4$.

The disk's inner edge $r_\inn$ is localized where the outwards torque of the $(2,1)$ OLR compensates the disk's inwards viscous torque. To calculate $r_{\inn}$, we use Eq.~10 of \citet{2013A&A...551A..50D}, which reads\footnote{Typos were present in the original expression : $r_\inn$ should be expressed in units of $a$ and the term $\sqrt{e}$ put outside the logarithm.}
\begin{equation}
  \frac{r_{\inn}}{a}=1.7+\frac{3}{8}\sqrt{e}\log(\mathcal{R}),
\label{rin}
\end{equation}
where $\mathcal{R}=(\alpha\gamma^{2})^{-1}$ is the Reynold's number, $\alpha$ is the viscosity parameter of \citet{1973A&A....24..337S}, and $\gamma=H/r_{\inn}$ with $H$ the disk's pressure scale height.

Our expression for $\dot{e}_\res$ contrasts with those of \cite{2013A&A...551A..50D}, \cite{vos2015} and \cite{oomen2020}, where in their formulation $\dot{e}_\res$ depends on the total angular momentum of the disk and on the viscous timescale. They used an approximation for $\dot{e}_\res$ proposed by \cite{lubow1996} that is valid when the disk structure has relaxed and is close to a steady state, which may be appropriate for a CB disk in post-AGB systems. The fact that in our model, matter is continuously fed into the disk, often faster than the viscous timescale, implies that the disk may not be in a steady state.

\subsection{The surface density at the OLR}
\label{sec:surface_density}

The value of the surface density at the resonance $\Sigma(r_\res)$ can be calculated by solving the time-dependent disk evolution equations, with appropriate boundary conditions \citep[see, e.g.][]{1991MNRAS.248..754P}.
The modeling of the disk's structure in such detail is beyond the scope of the current investigation.
Instead, we use a simple model to simulate the evolution of $\Sigma(r_\res)$:
\begin{equation}
  \frac{\partial{\Sigma}(r_\res,t)}{\partial t} = \xi\; \frac{\left|\dot{M}(t)\right|} {4 \pi a^2}-\frac{\Sigma(r_\res,t)}{\tau_{\visc}},
\label{dsig_dt}
\end{equation}
where $\dot{M}(t)<0$ is the systemic mass loss rate.
The first term on the right-hand side describes the injection of matter at the inner edge of the disk in a shell of radius $\Delta r \propto a$.
The parameter $\xi$ describes both the fraction of mass lost by the binary that is used to feed the disk, and the unknown distribution of matter near the disk edge. 
As will be seen in Sect.~\ref{sec:xi}, typical values for $\xi$ range between  $10^{-4}$ and $5\times 10^{-4}$. For smaller values of $\xi$  the effect of the CB disk is negligible, {while for larger ones, the density at the inner rim ($\Sigma_\inn$) is higher and leads to disk masses above $10^{-2}$\msun, in contradiction with current estimates in post-AGB stars  \citep{bujarrabal2013,oomen2020,bollen2022}.}.
The second term expresses the spreading of the disk and the associated decrease in $\Sigma(r_\res)$ by viscous torques that transport mass and angular momentum outward. The viscous timescale \citep{2001ApJ...548..900S}
\begin{equation}
  \tau_{\visc}= \frac{4r_{\inn}^{2}}{3\nu_{\inn}} =
  \frac{2}{3\pi\alpha\gamma^{2}}\left(\frac{r_{\inn}}{a}\right)^{\frac{3}{2}}P_\orb
\label{tau_visc}
\end{equation}
is evaluated at the disk's inner edge $r_\inn$. Here, $\alpha$ is the \cite{1973A&A....24..337S} disk parameter, $\nu_{\inn}$ the viscosity at $r_{\inn}$ given by
\begin{equation}
  \nu_{\inn} = \alpha c_{s}H = \alpha\gamma^{2}\sqrt{GMr_{\inn}},
\label{nu}
\end{equation}
and $c_{\mathrm{s}}$ is the sound speed. Integration of Eq.~\ref{dsig_dt} yields
\begin{equation}
  \Sigma(r_\res,t+\Delta t)=\Sigma(r_\res,t)e^{-\Delta t / \tau_{\visc}} + \xi \frac{\left|\dot{M}\right|}{4\pi a^2}\tau_{\visc} \left(1-e^{-\Delta t / \tau_{\visc}}\right)
\label{sigma_t}
\end{equation}
where $\Delta t$ is the time-step.
The disk temperature was estimated by \citet{2014A&A...568A..12H} using radiative-transfer calculations on a parametric CB disk structure. Their calculations indicate a temperature at the inner edge of $\approx 1600$~K,  declining with distance. Since this is below the ionization temperature of hydrogen, we assume $\alpha=0.01$ in Eq.~\ref{nu}, appropriate for cool disks where the magneto-rotational instability, thought to drive the viscosity mechanism, is suppressed \citep{1996ApJ...457..355G}. Additionally, studies of CB disk formation around cataclysmic variables \citep{2002ApJ...569..395D} suggest that $\gamma\lesssim 0.05$. We therefore take $\gamma=0.05$ and find $t_\visc \approx 10^5$ yr for $P_\orb=1000$~d, which gives a sensible disk lifetime.

\subsection{Stellar models}
\label{sec:stellarmodels}
We examine the evolution of two representative binaries consisting of solar metallicity stars with initial masses of 1.2+0.8\msun{} and 2.0+1.0\msun. We did not consider more massive primaries ($M_i\ge3$~\msun) because, as recently analyzed by \cite{Drisya}, the s-process abundances in Ba stars do not comply with the expected nucleosynthesis arising from the \chem{22}Ne($\alpha$,n) neutron source operating in these more massive AGB stars. The first system (1.2+0.8) considers the least massive primary for the star to evolve off the AGB within the Galactic lifetime. These lowest-mass primaries are also those that reach the smallest radius at the AGB tip and thus may help populate the short-period range of Ba stars.

The simulations are carried out using the stellar evolution code {\sf BINSTAR} \citep{2013A&A...550A.100S}. In the simulations, we adopt a solar composition ($Z=0.02$), according to \citet{1996ASPC...99..117G}. Convection is treated using the standard mixing length theory, with the mixing length parameter $\alpha_{\mathrm{MLT}}=1.75$, which results from fits of solar models. The input stellar physics is the same as described in \citet{Siess2008}.  The mass loss is prescribed according to \cite{SchroderCuntz07} up to the end of core helium burning and then during the AGB phase, we switch to the \citet{1993ApJ...413..641V} prescription. Radiative opacities are taken from \cite{IglesiasRogers96} above 8000\,K and from \cite{Ferguson05} at lower temperatures. The nuclear network includes 182 reactions coupling 55 species from H to Cl as described in \cite{Siess2008}. We also treat overshooting at the base of the convective envelope as a diffusive process, following \citet{2000A&A...360..952H} with $f_{\mathrm{over}}=0.01$. The evolution starts when the stars are on the zero-age main sequence (ZAMS) with the stellar spins synchronized with the orbital period.

The simulated 2.0+1.0\msun\ binaries have initial periods $P_i = \{4000,\ 6000,\ 8000,\ 12000, \ 16000\}$ days and initial eccentricities $e_i = \{0.01,\ 0.1,\ 0.3\}$. For 1.2+0.8\msun\ binaries, we compute additional models for $P_i =2000$ and 3000~d and the same eccentricities. As will be seen in Sect.~\ref{sec:results}, binaries with $P_\orb < 2000$~d completely circularize during their evolution, and above $P_\orb > 16000$~d WRLOF is not initiated (Sect.~\ref{sec:progenitors}). These systems are of limited interest and not considered in this study.  The effects of the wind speed and CB disk are evaluated by varying $v_\wind$ and $\xi$, whose values are summarized in Table~\ref{tab_param}. The evolution of the systems was followed until the end of the AGB phase, when convergence problems arose. At this point, less than $0.05\,\Msun$ remained in the envelope.
\begin{table}
  \begin{center}
    \caption{Wind and CB parameters used in our calculations.}
     \label{tab_param}
  \begin{tabular}{c c c c c c}
    \hline
    Model & $v_{\wind}$ (\kms) & $\xi$  \\
    \hline
    AV6     &  6                 & $5\times 10^{-4}$ \\
    AV10    &  10                & $5\times 10^{-4}$ \\
    AV15    &  15                & $5\times 10^{-4}$ \\
    BV6     &  6                 & $ 10^{-4}$ \\
    BV10    &  10                & $ 10^{-4}$ \\
    BV15    &  15                & $ 10^{-4}$ \\
    \hline
 \end{tabular}
\tablefoot{The parameter $\xi$ denotes the fraction of the mass lost by the binary used to feed the disk.}
\end{center}
\end{table}

\section{Conditions for WRLOF}
\label{sec:progenitors}

Using Kepler's third law and the \cite{1983ApJ...268..368E} formula for the Roche radius, the condition for WRLOF $R_\dust/R_{\Lone} = 1$ evaluated at periastron when it is most favorable, can be expressed in terms of the orbital period as
\begin{equation}
    P = (1-e)^{-3/2}\left(\frac{R_\dust}{f(q)}\right)^{\frac{3}{2}} \frac{2\pi}{ \sqrt{G(M_{1}+M_2)}},
    \label{eq:PorbCondition}
\end{equation}
where $R_{\Lone} \equiv a\;(1-e)\;f(q)$, and $f(q) = 0.49 q^{2/3} / (0.6q^{2/3} + \mathrm{ln}(1+q^{1/3}))$.
The orbital period, eccentricity and masses appearing in the previous equation do not correspond to the initial values because mass and angular momentum have been removed from the system by stellar winds.
In the absence of WRLOF, wind mass loss likely proceeds in the Jeans mode and the difference between the initial ($P_0$) and current period ($P$) at the onset of the WRLOF can be estimated using Eq.~\ref{eq:PeriodEvolution}.
Given the long periods of our systems, tides have a negligible effect prior to the onset of WRLOF, so the eccentricity can be considered constant, i.e. $e=e_0$.
Together, Eqs.~(\ref{eq:PeriodEvolution}) and (\ref{eq:PorbCondition}) give the initial period for the onset of WRLOF
\begin{equation}
  P_0^\WRLOF = (1-e_0)^{-3/2}\left(\frac{R_\dust}{f(q)}\right)^{\frac{3}{2}} \frac{2\pi}{\sqrt{G(M_{1}+M_2)}}  \left(\frac{1+q}{1+q_0}\right)^2.
  \label{eq:PorbCondition1}
\end{equation}
The dust condensation radius $R_\dust$  is given by Eq.~\ref{eq:Rdust} where in this equation the radius, effective temperature and mass of the primary star are taken from single stellar evolution models computed with the {\sf BINSTAR} code.
We further consider an O-rich dust with $T_\dust = 1000$~K and a circular orbit ($e_0=0$).

\begin{figure}

\includegraphics[width=\columnwidth]{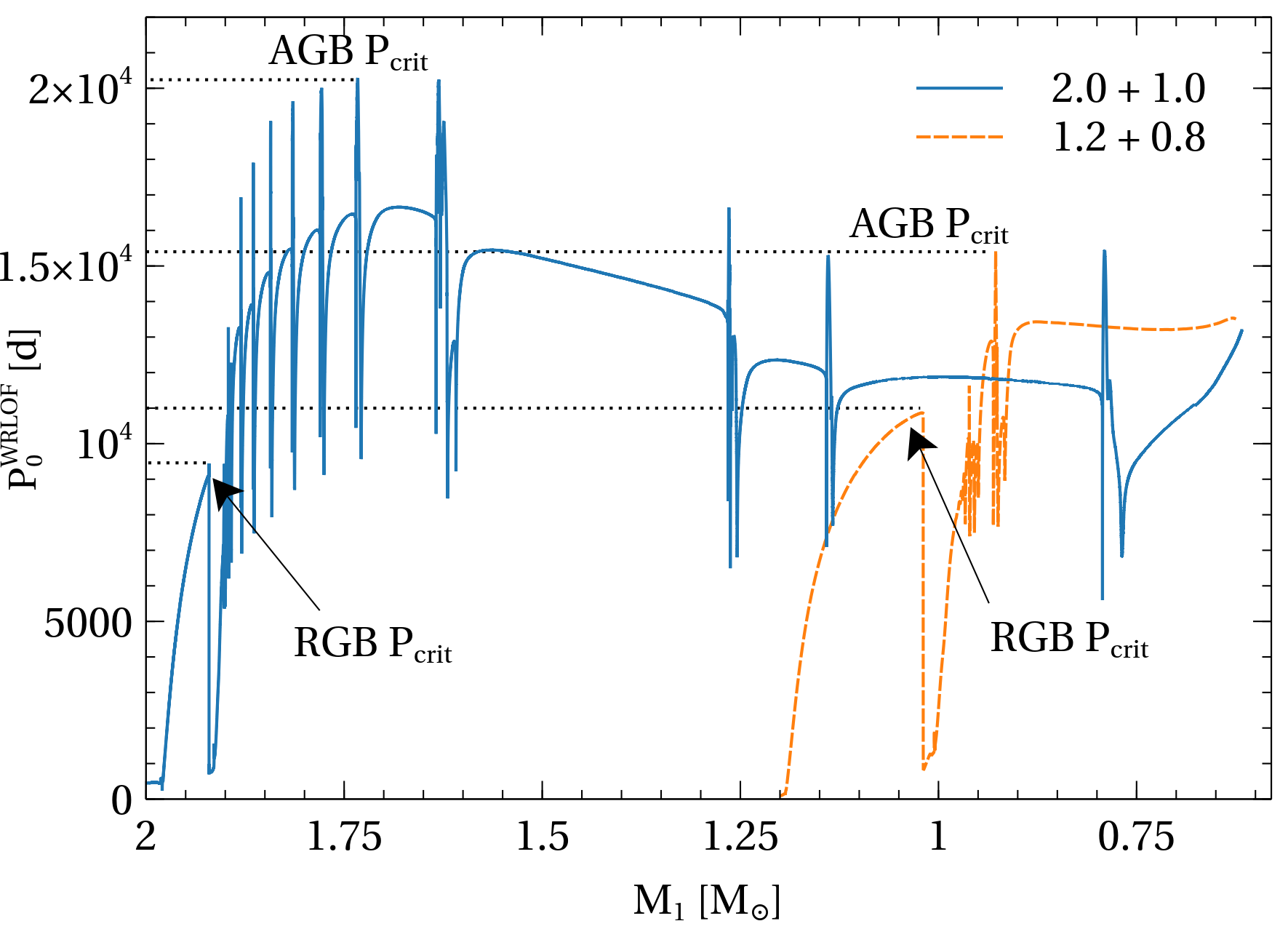}
\caption{Evolution of the initial period $P_0^\WRLOF$ for WRLOF as a function of the mass of the primary (Eq.~\ref{eq:PorbCondition1}) for a 2.0+1.0\msun\ (solid blue line) and a 1.2+0.8\msun\ (dashed orange line) binary. The spikes in the curves are associated with the development of thermal pulses. The black dotted lines indicate the critical initial period of circular binaries $P_\crit$ below which WRLOF first starts during the RGB or the AGB phase.}
    \label{fig:Pmax}
\end{figure}

Figure~\ref{fig:Pmax} shows the evolution of $P_0^\WRLOF$ as a function of the primary mass for the two systems considered in this study. The critical initial periods $P_\crit$ below which a circular 2.0+1.0\msun\ binary initiates WRLOF during the red giant branch (RGB) and AGB phases are respectively 9700~d and 20000~d (indicated by the horizontal dotted lines labeled RGB and AGB $P_\crit$ in Fig.~\ref{fig:Pmax}). For a circular 1.2+0.8\msun\ binary, the critical periods are shorter because of the smaller radius reached by lower-mass stars and ranges around 11000~d and 15400~d for the RGB and AGB phases respectively. During the AGB phase, WRLOF may be activated episodically during the short-lived thermal pulses (for the 2.0+1.0\msun{} system, this situation would appear if  $16500$~d $\lesssim P_0^\WRLOF \lesssim 20000$~d). However, due to the small amount of mass lost during these episodes, the orbital evolution is not significantly impacted, as shown by our tests.

\begin{figure}
\includegraphics[width=\columnwidth]{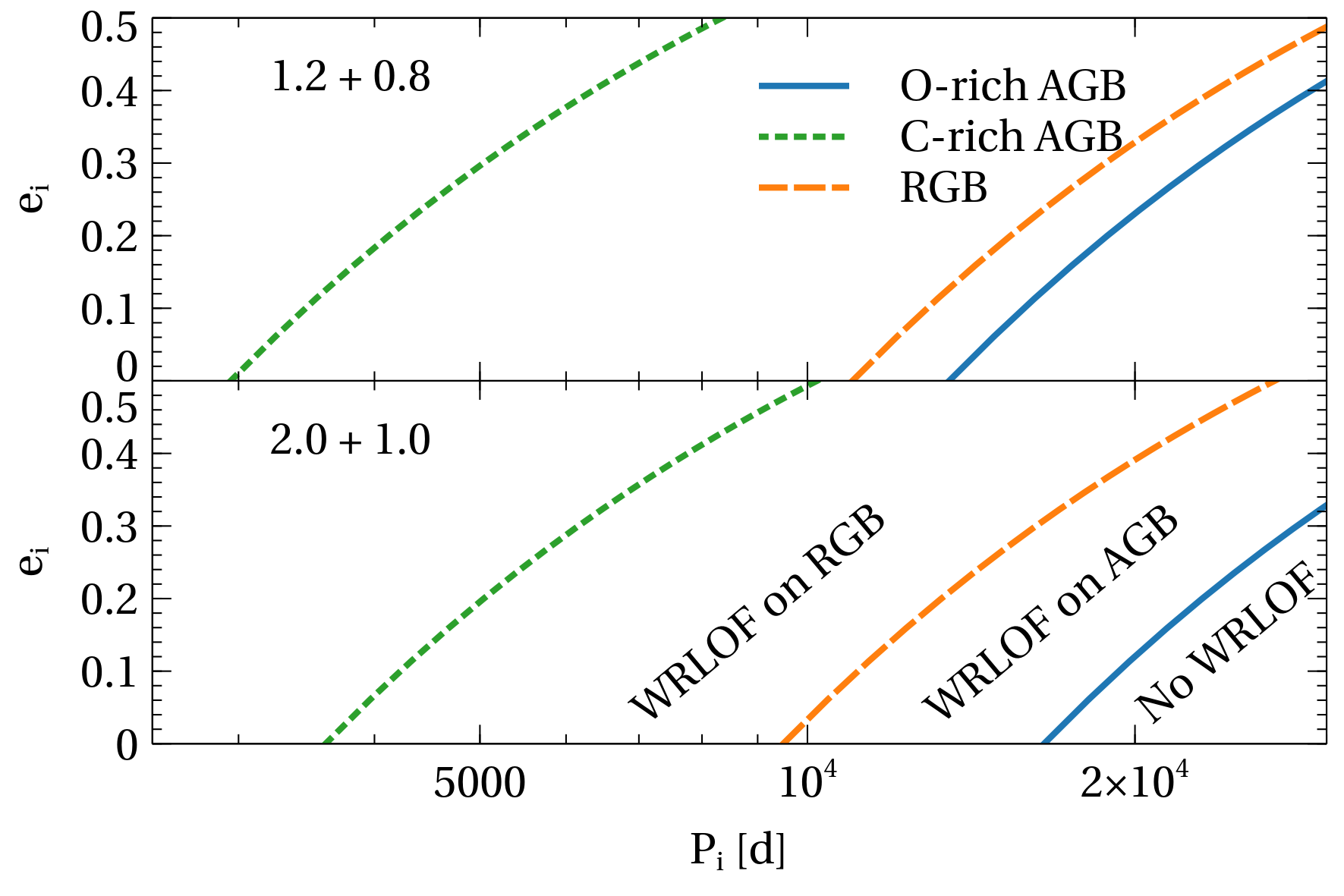}
\caption{Regions in the \elogP{} diagram where WRLOF is active. If the initial period $P_i$ of a system lies to the left of the curve, a binary undergoes WRLOF during the RGB (orange dashed line), the AGB (solid blue line), or the AGB with C-rich dust (i.e. $T_\dust{} =1500$~K, green short-dashed line). Top and bottom panels refer to 1.2+0.8\msun{} and 2.0+1.0\msun{} binaries, respectively. The orange dashed curve and solid blue curve are computed using $T_\dust{}=1000$~K (i.e. O-rich dust).}
\label{fig:Pmax-e}
\end{figure}

The dependence of the critical period for the onset of WRLOF on the RGB or AGB on the eccentricity and $T_\dust$ can be estimated from Eq.~\ref{eq:PorbCondition1} and is illustrated in  Fig.~\ref{fig:Pmax-e}.
With increasing eccentricity, binaries are more likely to experience WRLOF at any given period because of the reduction of the  periastron distance. As a consequence the critical period to start WRLOF at a specific phase increases.
For example, in the $2.0+1.0\Msun$ system, $P_\crit$ is $70\%$ longer for $e=0.3$, compared to the circular case.

Combining Eqs.~(\ref{eq:Rdust}) and (\ref{eq:PorbCondition1}) implies $P_i \propto T_{\dust}^{-15/4}$. This strong dependence on the dust condensation temperature implies that the critical period for C-rich stars will be a factor of five smaller than for O-rich stars.
The consequences of this composition change are quite significant.
If the conditions for the occurrence of WRLOF in a system involving an O-rich AGB star are initially met, the fact that the AGB star becomes C-rich, brings $R_\dust$ within its AGB Roche lobe and WRLOF stops. To have WRLOF during the C-rich AGB phase, the systems require shorter initial periods that may not be compatible with the absence of RLOF.

\section{Results}
\label{sec:results}

Before entering the detailed analysis of our models, we first describe our reference data sample of Ba stars that we use for comparison.

\subsection{The observational data}
\label{sec:sample}

\begin{figure}
\includegraphics[width=\columnwidth]{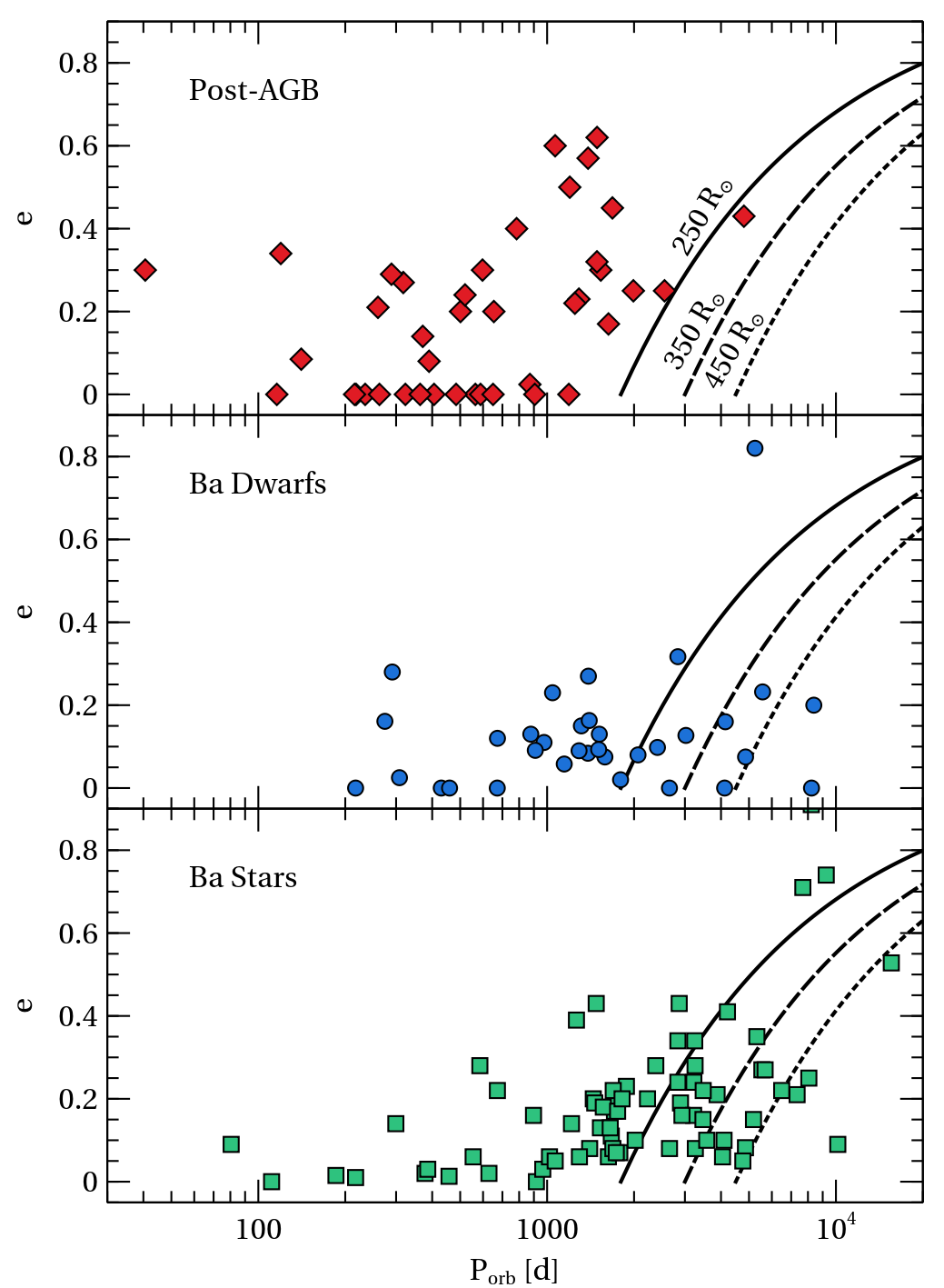}
\caption{Distribution in the \elogP{} diagram of the observed  post-AGB binaries  (top), Ba dwarfs (middle), and Ba giants (lower panel). The different lines delineate the orbital parameters for which a primary of radius 250\rsun{} (solid line), 350\rsun{} (dashed line) and 450\rsun{} (dotted line) fills its Roche lobe. These radii correspond to the maximum radii typically attained by solar-metallicity AGB stars of initial masses 1.2, 2 and 3\msun, respectively. Both components are assumed to have 1\msun{} at the time of the RLOF, although the exact masses do not  meaningfully affect the lines. As expected, lower primary-mass systems can reach shorter periods without RLOF. }
\label{fig:obs}
\end{figure}

The distribution of post-AGB stars, Ba dwarfs and Ba giants in the \elogP{}  diagram is shown in Fig.~\ref{fig:obs}. The data come from \cite{jorissen2016,jorissen2019}, \cite{swaelmen2017}, \cite{escorza2019} and references therein. The post-AGB sample is complete up to a period of $\approx 2000$\,d while, for dwarf and giant Ba stars this limit is pushed to  $P \approx 10^4$\,d.

The distributions (Fig.~\ref{fig:obs}) show a deficit of post-AGB and Ba dwarfs at large periods (above $P \gtrsim 2000$\,d) compared to the distribution of Ba stars which extends up to 30000~days. The reason for this difference can be attributed to the incompleteness of the sample for large periods but also to biases. In particular, Ba dwarfs, being fainter, are more difficult to observe, and on the main sequence, intermediate-mass Ba-dwarfs appear as A-type stars which are faster rotators, often variable (e.g., $\delta$ Scuti, $\gamma$ doradus and roAp stars), and known to show  chemical anomalies attributed to mixing processes (e.g., Am and Ap stars). Under these conditions, it may be difficult to classify an A-type star in the Ba-dwarf family.

It is important to stress that the post-AGB sample is likely contaminated by post-RGB systems. \cite{2014MNRAS.439.2211K} found that approximately two thirds of their Small Magellanic Cloud post-AGB binaries were in fact post-RGB systems.  Therefore comparisons between the stellar models and the post-AGB or Ba dwarfs will only be meaningful for relatively long-period systems.

As far as eccentricity is concerned, it seems that between the post-AGB and Ba dwarf stages, the overall eccentricity of the systems has decreased. This could be attributed to tidal effects circularizing the orbit on the long lifetime of the Ba dwarf provided the two classes are evolutionary connected. The evolution between the Ba dwarfs and the Ba giants has been investigated by \cite{Escorza2020} using the {\sf BINSTAR} code. Adopting as initial conditions the orbital and eccentricity distributions of Ba and CH dwarfs, these authors evolved those systems and were able to reproduce satisfactorily the \elogP{}  distribution of Ba giants.

Another limitation of the modeling comes from the fact that our binary code cannot handle unstable mass transfer, when the convective envelope of the AGB star overfills its Roche lobe. The outcome of such an evolution is uncertain and depends on various parameters, including the mass ratio at the beginning of mass transfer and on the dynamical response of the AGB star. It is generally believed that the expansion of the AGB star in response to mass loss leads to a common envelope phase during which a fraction of the orbital energy is transferred to the envelope, powering its ejection. During this process, the orbital parameters are likely to be affected but the magnitude of the changes are rather hard to estimate.

From single-star models computed with the {\sf BINSTAR} code \citep{2013A&A...550A.100S}, we find that in the mass range $1.2 \le M_1/\Msun \le 3$, the maximum radius at the AGB tip, $R_{\mathrm{AGB-tip}}$, varies between 250 and 450\rsun{}. Therefore, in Fig.~\ref{fig:obs}, systems that are
located in the \elogP{}  diagram to the left of the critical line defined by $R_{\Lone} = \min(R_\mathrm{AGB,tip}) \approx 250\Rsun$ will overfill their Roche lobe either on the RGB or AGB. The location of this critical line slightly depends on the mass ratio and, for circular systems, corresponds to $P \approx 2000$\,d. Therefore, our comparisons will be limited to those systems with $P\ga 2000$\,d.

\subsection{Analysis of a representative model}
\label{sec:detailed_system}

\begin{figure*}[t]
 \begin{center}
    \includegraphics[width=18cm]{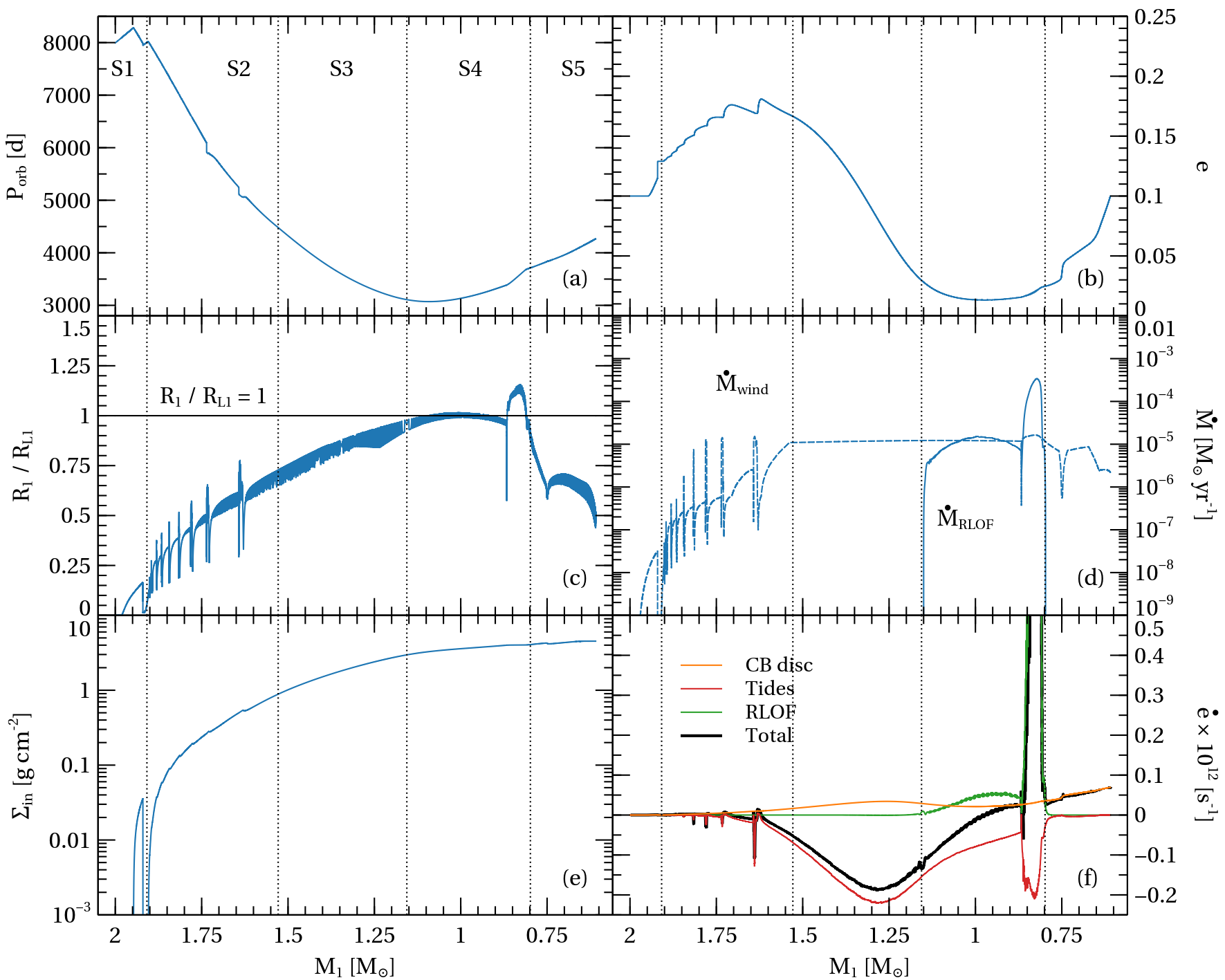}

    \caption{Evolution of a 2.0+1.0\msun{} binary with initial period $P_i=8000$ d and eccentricity $e_i = 0.1$ using the parameter set AV6. We present, as a function of the primary mass $M_1$,
    ($a$) the orbital period, $P_\orb$;
    ($b$) the eccentricity $e$;
    ($c$) the Roche-filling factor $R_1/R_{\Lone}$;
    ($d$) the primary mass loss rates from the wind $\dot{M}_\wind$ (dashed curve) and RLOF $\dot{M}_\RLOF$ (solid curve);
    ($e$) the surface density at the CB disk's inner rim, $\Sigma_\inn$;
    ($f$) the change in the orbital eccentricity $\dot{e}$  (black curve), which is decomposed into contributions from the CB disk (orange), tides (red), RLOF (green).
    The evolution of the system is divided into different stages, labeled S1-S5, and separated by the vertical black dotted lines.}
    \label{var_e_P}
    \end{center}
\end{figure*}

In this section, we study the orbital evolution of a 2.0+1.0\msun{} binary with $P_i =8000$~d, $e_i=0.1$, computed using the parameter set AV6 in Table \ref{tab_param}. The analysis of this representative model illustrates the various processes at play, providing a foundation for understanding the evolutionary tracks discussed later.
The evolution of this system is presented in Fig.~\ref{var_e_P}  and can be divided into five stages, labeled S1–S5. Note that not all models proceed through all these stages. For example, if a binary never undergoes RLOF, it may skip the S4 stage.

\paragraph{S1: Main sequence, RGB and core He-burning phase (epoch mass $M_{1} \gtrsim 1.9$\msun).}
The changes in orbital parameters during the main sequence and early RGB ($M_{1} \gtrsim 1.95$\msun) are small (panels $a$ and $b$).
At this stage, the dust condensation radius is within the primary's Roche lobe ($\RdRL <1 $), preventing WRLOF and the formation of a CB disk. The isotropic winds lead to a slight increase in the period, as expected from  Eq.~\ref{eq:PeriodEvolution}.  On the RGB, both $R$ and $R_\dust$ increase significantly and eventually, $\RdRL > 1$, briefly triggering WRLOF at $M_1 = 1.95$\msun. In this regime, the winds extract a larger amount of angular momentum (Eq.~\ref{eq:Jorbdot_WRLOF}), the orbit tightens and $P_\orb$ returns close to its initial value.
By the end of the RGB phase, the CB disk surface density $\Sigma_\inn$ reaches 0.03~\gcm{}  (panel $e$). When helium ignites at $M_1 = 1.9$\msun, the star contracts rapidly within its Roche potential, considerably reducing  the wind mass loss rate and the effects of tides which become inefficient (panel $f$). When resonant interaction with the CB operates, it produces a small increase in eccentricity from 0.1 to 0.13 until the disk dissipates, and $\Sigma_{\inn}$ vanishes (panel $e$) after a few viscous timescales (with $\tau_{\visc}$ of the order of  $\sim10^5$~yr; Eq.~\ref{tau_visc}).
The orbital evolution of the system up to the end of the core He-burning phase is thus characterized by a small increase in eccentricity  by the CB disk and a small change in the orbital period given the small amount of mass lost during S1 ($\sim 0.1$\msun).

\paragraph{S2: Disk-dominated phase (epoch mass $1.9 > M_1 / \Msun> 1.55$).}
During phase S2, the dust condensation radius moves again past the Roche lobe radius and $\RdRL$ will remain larger than unity until the end of the simulation. The activation of WRLOF will allow the CB disk to reform (panel $e$) and produce a strong orbital contraction due to the systemic wind mass loss. Interestingly,  the CB disk has a very weak effect on the separation.  Mass feeding into the disk increases $\Sigma_\inn$ (panel $e$) and leads to efficient eccentricity pumping ($\dot{e}_\CB$, orange curve in panel $f$). But, as the star expands on the thermally pulsing asymptotic giant branch (TP-AGB) and the separation keeps shrinking, tides become stronger ($\dot{e}_\tides$, red curve).

Thermal pulses can be identified as spikes separating the interpulse phases in panels $c$ and $d$ displaying to the overfilling factor and the wind mass loss rate, respectively. During the relaxation of the structure following the energy release by the pulse, the star contracts, reducing the efficiency of the tides and the mass loss rate. This allows the CB disk to act almost unconstrained,  producing small jumps in $e$ (panel $b$).

The TP-AGB phase is thus characterized by a decrease in the orbital period (panel $a$), caused primarily by WRLOF systemic mass loss. The CB disk acts efficiently during this stage, resulting in a growth of the eccentricity (panel $b$). It is important to remember that in binaries with a shorter initial period, tides will be stronger and circularization may already begin in this phase.

\paragraph{S3: Tidally dominated phase (epoch mass $1.55 > M_1/ \Msun> 1.15$).}
Throughout this phase, tides continue to strengthen as the star expands and the orbital period decreases, resulting in a near circularization of the system ($e \approx 0.01$ at $M_1 = 1.15$\msun, panel $b$). The wind mass loss rate reaches its superwind value of $\approx 10^{-5}\Myr$, leading to the accretion  of 0.25\msun{} onto the companion while in the meantime the primary loses 0.85\msun. These numbers are in line with the $\sim 30\%$ accretion efficiency predicted by our model (Fig.\ref{fig:eta_parameter}, third and fourth panels). The shrinkage of the separation is mainly due to systemic wind mass loss, the mass transfer onto the secondary providing a small positive contribution to $\dot{a}$.

\paragraph{S4: RLOF mass-transfer phase (epoch mass $1.15 > M_1/ \Msun> 0.8$).}
The expanding AGB star eventually fills its Roche lobe at $M_1 \approx 1.15$\msun (panel $c$) and mass transfer starts.
It is important to stress that the companion is more massive than the AGB star ($M_2 = 1.25$\msun while $M_1 =1.1$\msun) and that RLOF begins in an eccentric orbit ($e=0.015$, panel $b$). This unusual situation leads to an overall increase in the separation, despite the tendency of tides to circularize the orbit.
The phase-dependent RLOF in the eccentric orbit also produces a modest increase in eccentricity (Eq.~\ref{eq:edot_exRLOF} and panel $f$).
The abrupt variation in $R_1/R_\Lone$ and $\dot{M}_\RLOF$ (panels $c$ and $d$) at $M_1 = 0.85$\msun{} is due to the development of a thermal pulse and occurs just before the termination of  RLOF at $M_1\approx 0.8$\msun.

\paragraph{S5: End of the AGB phase (epoch mass $0.8 > M_1> 0.6$\msun).}
During this final stage, the AGB star expels its remaining envelope through winds. As the convective envelope becomes thinner, tides become inefficient. The separation now increases due to wind accretion onto the companion, which has become stronger than the orbital shrinkage due to WRLOF.
Furthermore, the eccentric RLOF prevented the complete circularization of the binary, enabling efficient resonant interactions with the CB disk in this final stage (panel $f$).
This simulation ends at $M_1 = 0.6$\msun{} with an eccentricity of 0.1 and a period of 4300~d.

As for most stellar evolution codes, convergence difficulties arise at the tip of the AGB \citep[e.g.][]{Wagenhuber1994,Lau2012}, preventing the primary from evolving on the post-AGB and WD cooling tracks. Given the significant decline in the mass loss rate and the short duration of the  post-AGB phase, the impact of winds, tides, and RLOF on the orbital parameters is probably limited. However, the CB disk can still act on the eccentricity before dissipating, as will be discussed in Sect.~\ref{sec:evolution_Post-AGB}. 

\subsection{Evolution in the \elogP{}  diagram}
\label{orbit_evol}

\begin{figure}

\includegraphics[width=\columnwidth]{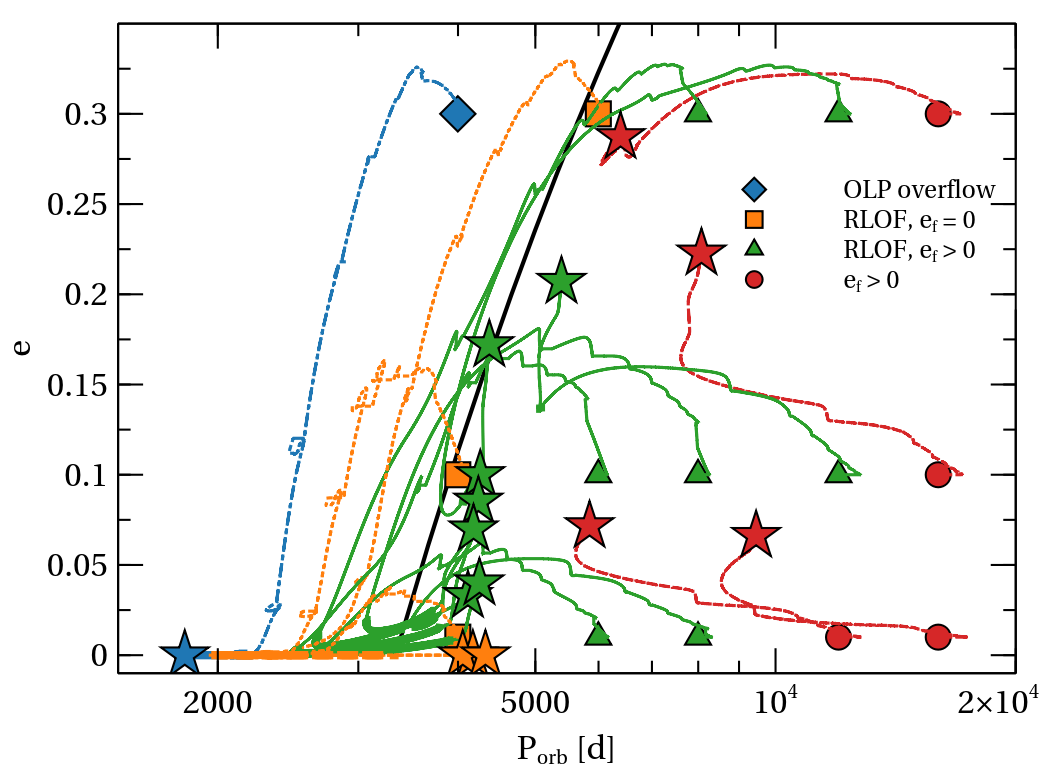}
\caption{Evolution in the \elogP{} diagram of the 2.0+1.0 binaries using parameter set AV6.  The color of the points refers to the outcome for the system at the end of the simulation. Red represents eccentric final orbits ($e_f>0$), green represents the same, but with the added condition that the binary undergoes RLOF. Orange corresponds to binaries that experienced RLOF, but ended up on a circular orbit. Star symbols show the final orbital parameters. The blue track identifies systems where the AGB radius exceeds the outer Lagrangian point (OLP), at which point,  marked by the blue star, the simulation is stopped. The orbital evolution of binaries  is illustrated with an evolutionary track, where the thickness of the line increases once the RLOF starts.
The solid black line corresponds to orbital parameters such that a 1\msun\ primary at the tip of the AGB (having a radius of 350\rsun) fills its Roche lobe, assuming a 1\msun{} companion.}
\label{fig:e_logP_tracks}
\end{figure}

The orbital evolution in the \elogP{} diagram of Fig.~\ref{fig:e_logP_tracks} can be understood from our previous analysis of the 8000~d+0.1 system. For systems with $v_\wind= 6$~\kms and orbital periods shorter than 20000~d, the activation of WRLOF during the RGB produces  a decrease of the orbital period (because $\eta > \eta_\crit$). The formation of the CB disk then allows the eccentricity to grow up to core He ignition. During the AGB, WRLOF resumes, the separation decreases and the eccentricity starts rising again until tides take over and start circularizing the system.

Eventually, the binaries that come sufficiently close can undergo RLOF, as depicted in Fig.~\ref{fig:e_logP_tracks} by a thicker curve. As long as the mass ratio $q$ stays above unity, the period and eccentricity decrease, but when $q<1$,  both $P_\orb$ and $e$ begin to increase again, producing a loop at the end of the evolutionary track.
The final orbital evolution follows three possible paths, depending on whether the final orbit is circular or eccentric and RLOF is stable or not. We consider that RLOF is unstable if the radius of the AGB exceeds its outer Lagrange point (OLP) determined using Eq.~4 of \cite{temmink2023}. In that case, the simulation is stopped and the result discarded.

The first scenario involves binaries with $P_i = 16000$~d, the 12000d+0.01, and 12000d+0.1 systems (red circles in Fig.~\ref{fig:e_logP_tracks}). These systems end their evolution on eccentric orbits and avoid RLOF. The final period ranges between 6000 and 9000~d and owing to the CB disk, the final eccentricities hit values between 0.2 and 0.3.

The second scenario includes the 12000d+0.3 binary, systems with $P_i = 8000$~d, and two with $P_i=6000$~d (green triangles in Fig.~\ref{fig:e_logP_tracks}). These binaries undergo stable RLOF in eccentric orbits and because of the phase-dependent mass transfer, the eccentricity is also pumped during RLOF. This process is effective in nearly circularized binaries such as the 8000~d+0.1 system.
However, with the reversal of the mass ratio ($q<1$), the separation will increase again and  prevent the system from reaching periods below the black line in Fig.~\ref{fig:e_logP_tracks}, which roughly corresponds to the critical limit for the onset of RLOF.

The final scenario (orange squares) accounts for binaries with initial periods shorter than 6000~d, the exact value depending on the eccentricity. Owing to the greater proximity of the stars, tidal forces quickly circularize the orbit. Consequently, when RLOF begins, the binary system cannot regain some eccentricity, as in the previous scenario.  The final period of  these circular systems ranges between 2000 and 5000~d.

We stress that at the end of the simulation, the CB disk may still be present and may further impact the orbital evolution. This effect will be assessed in Sect.\ref{sec:evolution_Post-AGB}.

\begin{figure*}[t]
  \begin{center}
    \includegraphics[width=15cm]{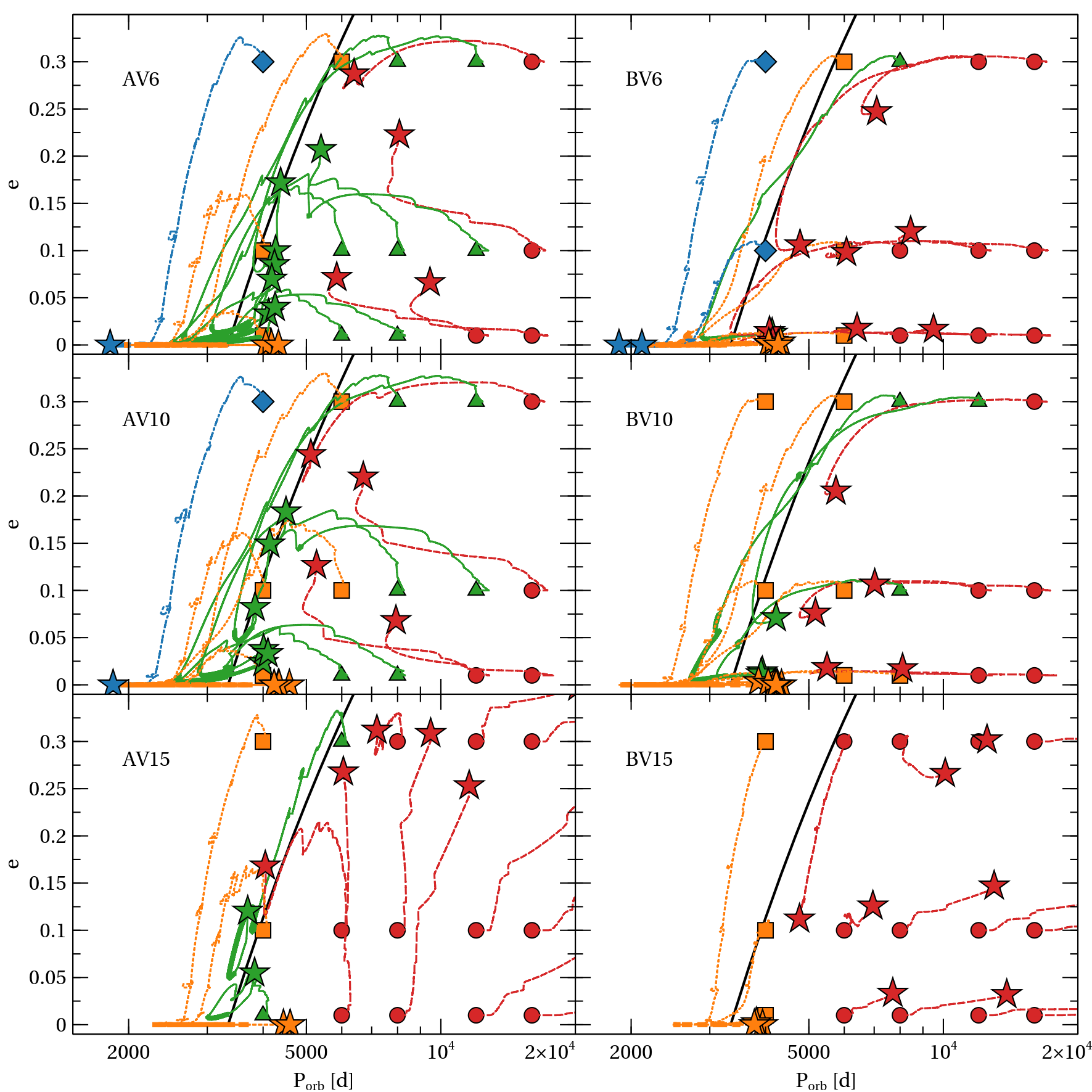}
    \caption{Same as Fig.~\ref{fig:e_logP_tracks} but comparing the impact of different wind speed $v_\wind = 6$, 10, and 15 km/s and with  $\xi = 5\times 10^{-4}$ (left column, sets AV6, AV10, AV15) and $\xi = 10^{-4}$ (right column, sets BV6, BV10 and BV15).}
    \label{fig:e_logP_models}
    \end{center}
\end{figure*}

\subsection{Effect of parameter variations}
\label{orbit_impact}

To investigate the effects of the wind velocity, $v_\wind$, and disk parameter, $\xi$ (with values as listed in Table~\ref{tab_param}), we computed additional models for the 2.0+1.0~\msun{} system but keeping the same grid in orbital parameters, i.e. $P_i = \{4000,\ 6000,\ 8000,\ 12000,\ 16000\}$~d, and $e_i= \{0.01,\ 0.1,\ 0.3\}$.

\subsubsection{Effect of changing the wind speed $v_\wind$}
\label{sec:vwind}

We first focus on models AV6, AV10, and AV15, which use the same value of $\xi=5\times 10^{-4}$ but differ in the selected wind speed, with $v_\wind = 6$, 10 and 15~\kms{} respectively. The results of these models are displayed in the left column of Fig.~\ref{fig:e_logP_models}.

Increasing the wind speed reduces the specific angular momentum carried by the ejected material ($\eta$, see Fig.~\ref{fig:eta_parameter}) as well as the accretion efficiency $\beta$ onto the companion. Models AV6 and AV10 present essentially the same evolutionary tracks in the \elogP{} plane. In the AV10 models, the wind extracts less specific angular momentum but more material escapes from the system (due to the lower accretion efficiency onto the companion). The combination of these two antagonistic effects results in similar angular momentum losses for models AV6 and AV10, and in the end, they lead to comparable evolution.

On the contrary, the AV15 models behave quite differently: only the 6000~d+0.3 and $P_i=4000$~d systems  experience RLOF and in the remaining models (red track), only the 6000d+0.1 and 8000d+0.3 binaries stay close to their initial periods whereas wider systems expand during their evolution. This can be explained from inspection of  Fig.~\ref{fig:eta_parameter} which shows that for $v_\wind =15$~\kms{} and $P_\orb \gtrsim 6000$~d, $\eta<\eta_\crit$ so the systemic wind mass loss will cause the separation to increase. 
Because AV15 systems remain in long orbital periods, tides have a negligible effect and the resonant interactions with the CB disk can efficiently pump the eccentricity up to 0.3 in systems with initial periods between  6000~d and 8000~d. In systems with larger separations ($P_i\geq 12000$~d), the increase in $\Sigma_\inn$ is lower (Eq.~\ref{dsig_dt}) and the effect of the disk weaker, resulting in a wider range of eccentricities, distributed between 0.1 and 0.35.

\subsubsection{Effect of surface disk density parameter $\xi$}
\label{sec:xi}

The models in the right column of Fig.~\ref{fig:e_logP_models} use the same values of $v_\wind$ as their left counterparts, but with $\xi = 10^{-4}$ ($\xi$ represents the fraction of the mass lost by the primary that feeds the disk). With decreasing values for $\xi$, the density at the disk inner rim $\Sigma_\in$ (Eq.~\ref{dsig_dt}) is smaller and the impact of the CB disk ($\dot e_\res$) is reduced.

In models BV6, BV10, and BV15, the effects of the disk are negligible and the results are essentially the same as for the same models computed with $\xi = 0$. In the \elogP{} diagram, models BV6 and BV10 evolve toward shorter orbital periods at nearly constant $e$, until tides become efficient and circularization proceeds. In the BV15 models, binaries with $P_i \geq 6000$~d expand under wind mass loss and in this case the eccentricity is not reduced. It can even slightly increase due to phase-dependent wind accretion.
The surface densities in these models are of the order of 0.1-1~\gcm{} (compared to 1-10~\gcm{} in models with $\xi = 5\times 10^{-4}$). This gives an estimate of the lower limit on $\Sigma_\inn$ that is needed for the effects of the CB disk to manifest during the ascent of the AGB.

The comparison between models A and B  (left column of Fig.~\ref{fig:e_logP_models}) shows that the occurrence of RLOF does not greatly depend on the value of $\xi$. However, due to the higher value of $\xi$, models A are more prone to start RLOF on an eccentric orbit, leading in the end to a more efficient eccentricity pumping.

\subsubsection{Effect of changing the tidal strength}

\cite{Nie2017} suggested that a reduction of the tidal circularization rate by up to two orders of magnitude is needed to explain the  existence of ellipsoidal red giants in eccentric binary systems.
Following \cite{Escorza2020}, we consider this effect in models BT01, where the $\dot{e}_{\tides}$ terms are reduced by a factor 10 ($F_\tide = 0.1$).

Our results indicate that binaries with $P_\orb = 16000$~d are too wide for tidal effects to play any role and their tracks are identical to the corresponding systems in model B. For shorter periods, the evolution initially proceeds as in models B, with a contraction of the orbit due to systemic wind mass loss and an increase in the eccentricity resulting from interactions with the CB disk. When mass transfer starts (around 3000~d), the eccentricity is higher than in model AV6 and the $e$-pumping by RLOF more efficient. When the mass ratio inverts, the orbit widens and the eccentricity can reach 0.6.
In conclusion, $F_\tide$ has a significant impact on the orbital parameters as it allows to reach higher eccentricities but it does not help in producing shorter period systems.

\subsection{The tidally enhanced wind scenario}
\label{impact_crap}

\begin{figure}
\includegraphics[width=\columnwidth]{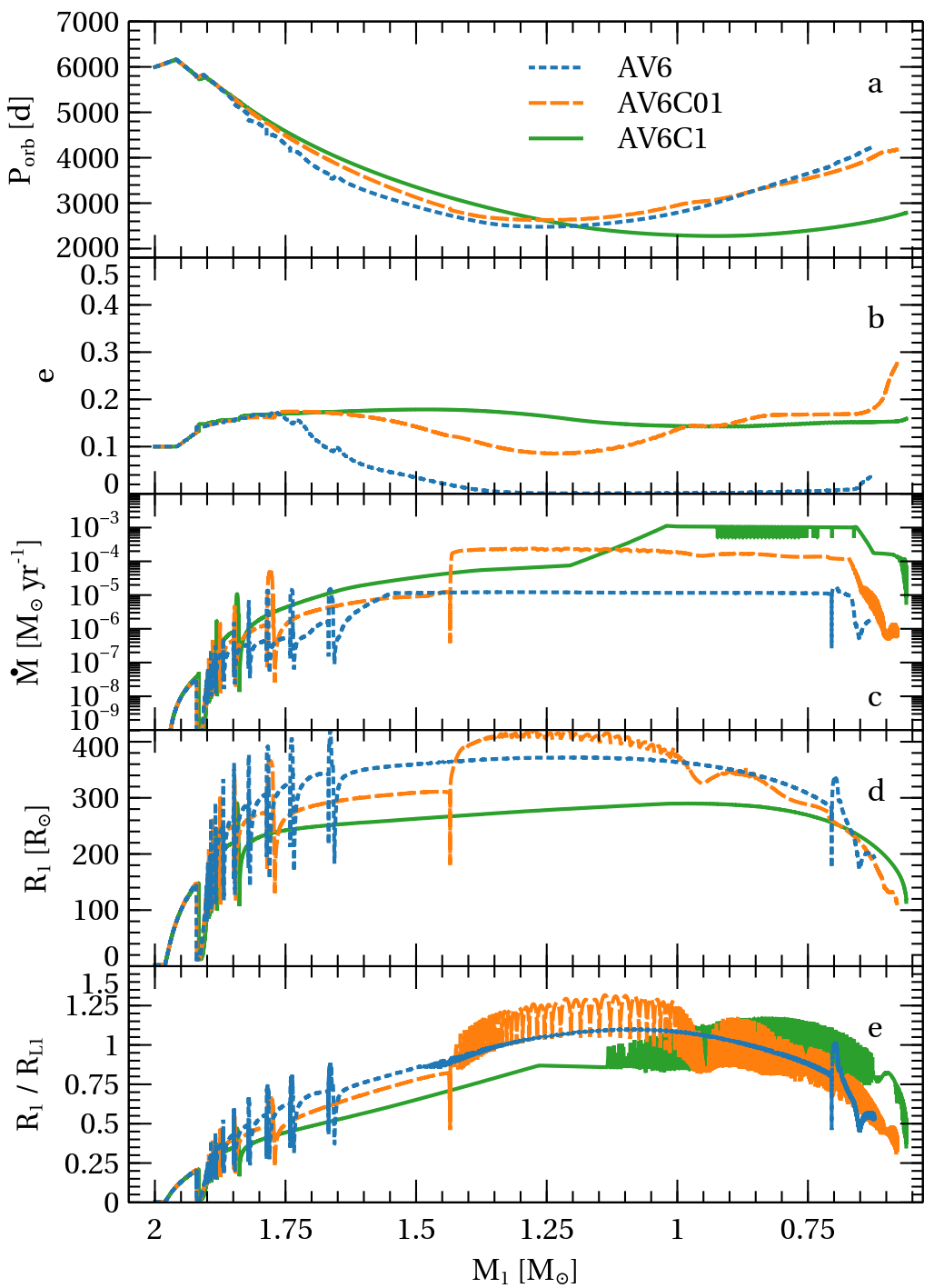}
\caption{Orbital period $P_\orb$ ($a$), eccentricity ($b$), primary mass loss rate ($c$), radius $R_1$ ($d$), and Roche filling factor $R_1 / R_{\Lone}$ ($e$) for the 2.0+1.0\msun{} system with $P_i$ = 6000~d, $e_i$ = 0.1 and different values of $B_\wind$: 0 (dotted blue line), $10^3$ (dashed orange  line), and $10^4$ (solid green line).}
\label{fig:CrapEvolutionComparison}
\end{figure}

To account for the observed systems with $P_\orb \lesssim 2000$~d, the evolution of the primary must be truncated before its reaches its maximum radius, thereby allowing the binary to shrink to shorter orbital periods before RLOF occurs. We investigate in this section if tidally enhanced wind loss could produce this effect.

\citet{ToutEggleton1988} suggested that the presence of the companion star could enhance the primary's wind mass loss rate. They propose a formulation for the wind loss rate, $\dot{M}_\wind$, given by
\begin{equation}
  \dot{M}_\wind=\dot{M}_{\wind}^{\mathrm{VW}}\times\left\{1+B_\wind
    \times\mathrm{min}\left[\left(\frac{R_{1}}{R_{\Lone}}\right)^{6},\frac{1}{2^6}\right]\right\},
\label{crap}
\end{equation}
where $\dot{M}_{\wind}^{\mathrm{VW}}$ is the AGB wind-loss rate \citep[][in our case]{1993ApJ...413..641V} and $B_\wind$ is a free parameter. The last term in square brackets accounts for a saturation when the star has filled more than one half of its Roche lobe. \citet{ToutEggleton1988} found that $B_\wind\approx 10^{4}$ could match the orbital parameters of the RS CVn binary Z Her, a value not dissimilar from that of \cite{2014A&A...565A..57S} ($B_\wind\approx  3.6\times 10^4$)  to account for the properties of IP Eri, a long-period eccentric binary with a helium WD.

Figure \ref{fig:CrapEvolutionComparison} shows the evolution of the  orbital period, eccentricity and primary-star radius for a 2.0+1.0\msun, 6000d+0.1 system using the parameters from model AV6, but with $B_\wind = 10^3$ (orange) and $B_\wind = 10^4$ (green). The standard model with $B_\wind = 0$ is shown in blue.

The tidally enhanced mass loss rate has two effects. First, it accelerates the AGB evolution of the primary and thus reduces the timescale during which tidal circularization and the interaction with the CB disk can operate. As a consequence, the eccentricity remains relatively constant (green and orange curves, panel $b$) contrary to the system computed without the wind enhancement which nearly circularizes (blue curve).
Second, by peeling-off the surface layers more efficiently, the enhanced mass loss limits the expansion of the star which reaches a maximum radius of 260\rsun{} for $B_\wind = 10^{4}$, compared to $350$\rsun{} in the standard ($B_\wind=0$) case (panel $d$). Consequently, in the $B_\wind = 10^{4}$ system, RLOF begins at a shorter period ($\sim 2200$~d) than in the standard case (at $P_\orb \approx 2400$~d, panel $a$).
During the mass-transfer episode, the $B_\wind = 10^{4}$ model experiences a modest increase in $P_\orb$, to 2600~d, contrarily to the other two cases, where $P_\orb$ increases to 4000~d.
This stems from the fact that in this model the rates of wind mass loss and RLOF mass loss are comparable, of the order of $\sim 10^{-3}\Myr$, so their effects on the orbit compensate. This situation, occurring for the largest value $B_\wind$,  is unusual because typically one of the two processes (wind mass loss or RLOF mass transfer) predominates. Therefore, the results for this model should be approached with caution, as the simple modeling applied here may not be physically sound.

The model with $B_\wind = 10^{3}$ (orange curves) has a special evolution in the sense that the  expansion of the primary (up to $400$\rsun, panel $d$) following the thermal pulse at $M_1 = 1.43$\msun{} leads to a quite large Roche-filling factor ($R_1 /R_\Lone = 1.25$, panel $e$). In this case, the RLOF mass-transfer rate remains larger ($5\times  10^{-3}\Myr$) than the wind mass loss rate ($10^{-4}\Myr$, panel $c$). Therefore, contrarily to the $B_\wind = 10^{4}$ case, the orbital period increases to 4000~d after RLOF, similarly to the standard case.

\begin{figure}
\includegraphics[width=\columnwidth]{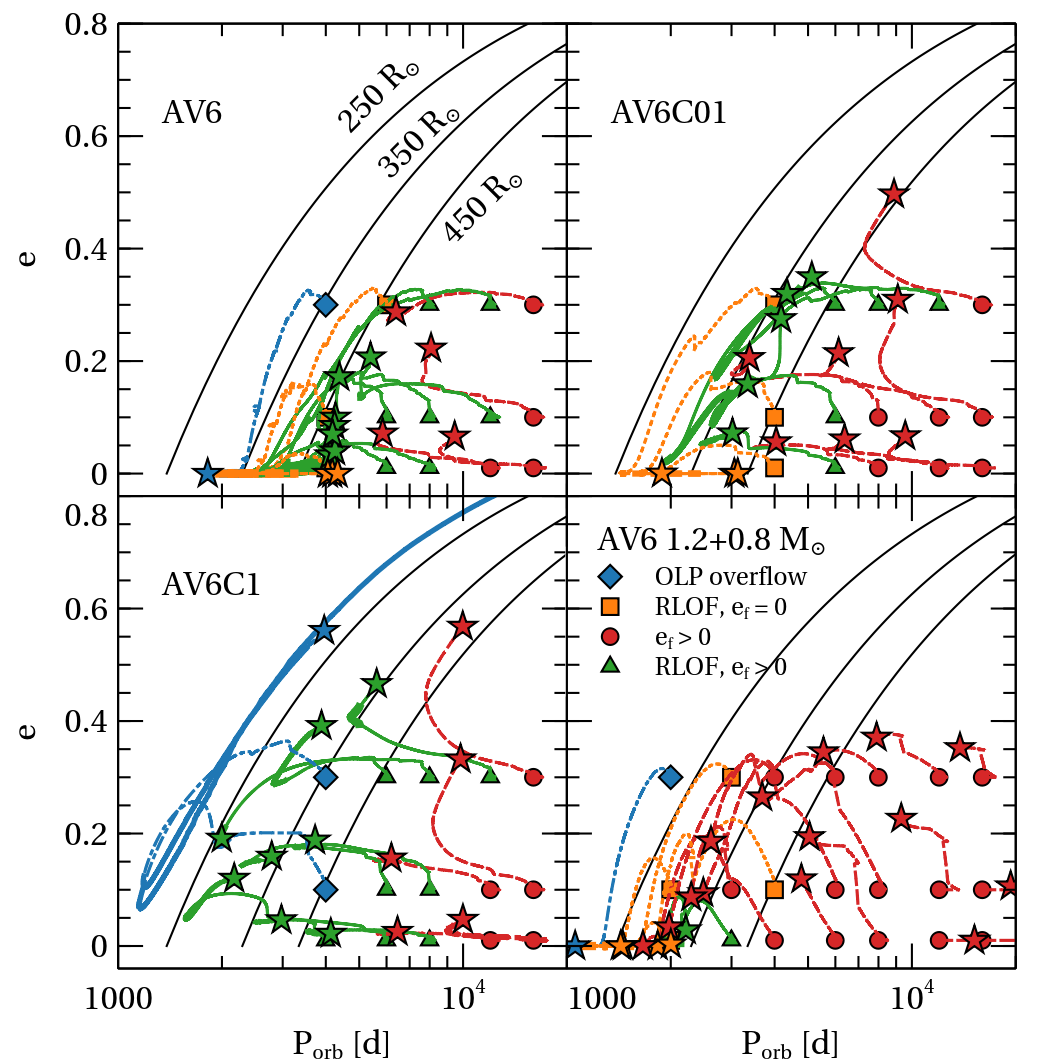}
\caption{Same as Fig.~\ref{fig:e_logP_tracks} but for the tidally enhanced wind simulations (parameter sets AV6, AV6C01, and AV6C1) and for the lower-mass system AV6, 1.2+0.8\msun, bottom right panel). For the massive systems, the three solid lines correspond to orbital parameters where a 250, 350, 450\rsun{} primary fills its Roche lobe, assuming a mass ratio of 1.}
\label{fig:e_logP_CRAP}
\end{figure}

We simulated a grid of binaries using the same parameters as for AV6, but with $B_\wind = 10^3$ (model AV6C01) and $B_\wind = 10^4$ (model AV6C1). The evolutionary tracks and final orbital parameters are shown in Fig.~\ref{fig:e_logP_CRAP}. When the tidally enhanced wind is present (panels AV6C01 and AV6C1), binaries with $P_\orb \lesssim 8000$~d are more likely to initiate RLOF on eccentric (green triangles) rather than circular (orange squares) orbits. This results from the shortening of the AGB phase which reduces the time during which tidal circularization can act. These binaries initiate RLOF in more eccentric orbits and end up with higher final eccentricities, up to 0.4 and 0.5 for $B_\wind = 10^{3}$ and $10^{4}$ models respectively, compared to a maximum of $\sim 0.3$ in  the AV6 case. Additionally, because the primaries do not grow as big and start RLOF in tighter orbits, they can attain shorter final orbital periods, as low as $3000$~d for $B_\wind = 10^{3}$ and $2000$~d for $B_\wind = 10^{4}$ compared to $4000$~d in the non-enhanced case. Remarkably, the region in the \elogP{} where the AV6C01 and AV6C1 models end their evolution coincides with the lines where a primary with a radius of 350\rsun\ and 250\rsun\ fills its Roche lobe, assuming at that time a mass ratio $q \sim 1$ (solid black line in Fig.~\ref{fig:e_logP_CRAP}). These radii agree well with the maximal radius attained by  AV6C01 and AV6C1 models.

\subsection{Non-conservative RLOF}

\begin{figure}
\includegraphics[width=\columnwidth]{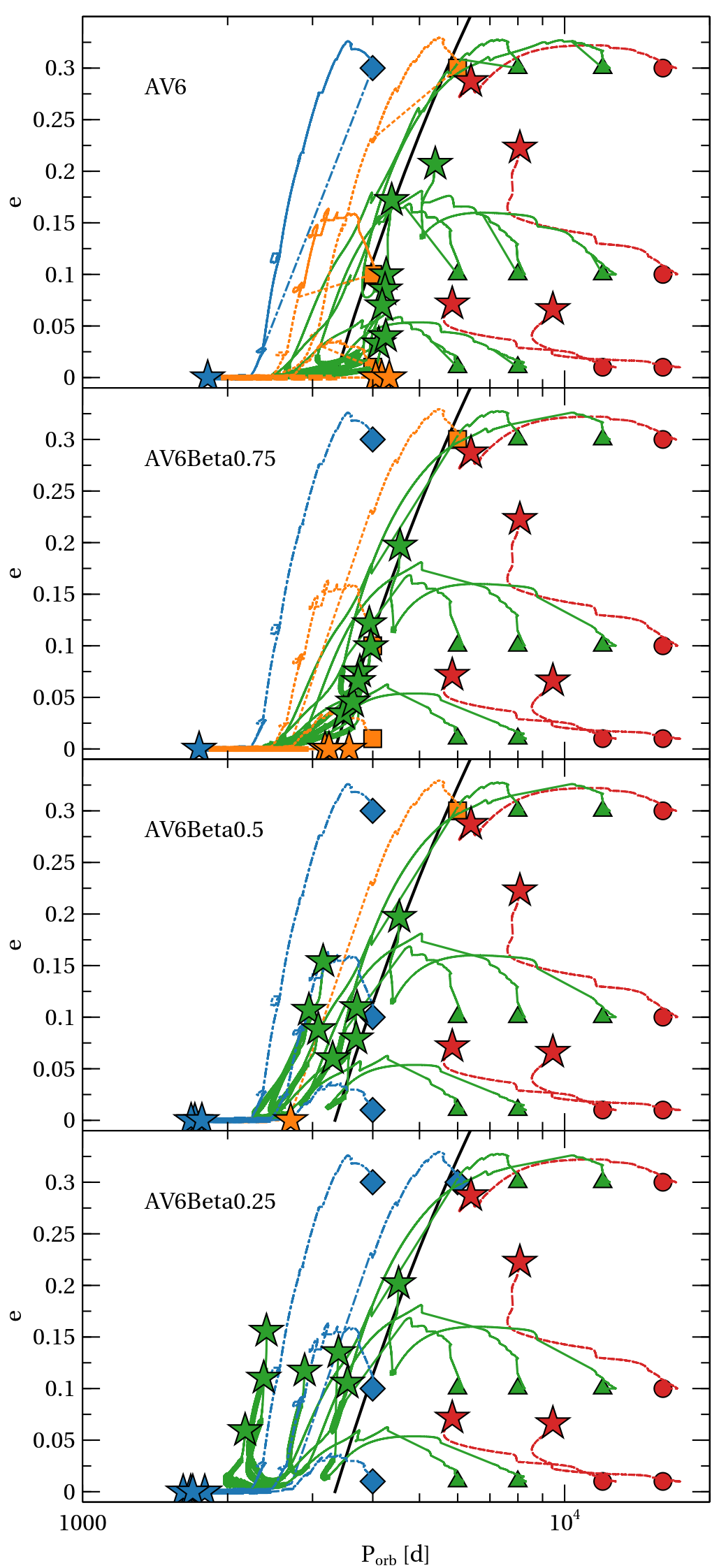}
\caption{Same as Fig.~\ref{fig:e_logP_tracks} but for non-conservative RLOF. The accretion efficiencies $\beta_\RLOF$ are indicated in the upper-left corner of each panel. Conservative mass transfer corresponds to $\beta_\RLOF = 1$ (i.e. AV6). The part of the tracks displayed with thick solid lines refers to ongoing RLOF. }
\label{fig:elogP_NC}
\end{figure}

During conservative RLOF, mass transfer from a lower-mass donor widens the orbit, preventing the binary from reaching orbital periods significantly shorter than the initial period at which RLOF began. In the case of non-conservative RLOF, if sufficient mass is lost with high specific angular momentum, the separation can keep shrinking even for a mass ratio less than one.

Non-conservative RLOF is modeled  using a constant accretion efficiency $\beta_\RLOF \leq 1$, such that the gainer accretes at a rate  $\dot{M}_2 = -\beta_\RLOF\; \dot{M}_1$.  When $\beta_\RLOF < 1$, mass is lost from the system at a rate $\dot{M} = (1 - \beta_\RLOF)\; \dot{M}_1$ and carries away  angular momentum at a rate given by  Eq.~\ref{eq:Jorbdot_WRLOF}. In this formulation we impose $\eta = 0.6$, corresponding to the maximum value of $\eta$ found in the WRLOF simulations of  \citet{saladino2019} for very low wind velocities.  We performed simulations for the 2+1\msun{} binary using the parameter set AV6, and considered three accretion efficiencies $\beta_\RLOF = \{0.25,\ 0.5,\ 0.75\}$.

Non conservatism only affects the final evolution of the models that cross the solid black line in Fig.~\ref{fig:elogP_NC} which represents the limit below which RLOF is expected to occur during the AGB phase.
Models with $\beta_\RLOF=0.75$ are quite similar to the conservative ones. The binaries that go through the RLOF phase end up with similar eccentricities but they reach slightly shorter period (between 3500~d and 4200~d compared to 4000~d and 5000~d when $\beta_\RLOF=1$).
The models with $\beta_\RLOF$ values of 0.5 and 0.25 reach periods as low as 3000~d and 2100~d, respectively. These results show that the degree of conservativeness during RLOF can play a significant role in determining the final orbital parameters. Nevertheless, the observed eccentric barium stars and post-AGB binaries with periods shorter than 2000~d remain unaccounted for.

\subsection{Influence of the initial binary masses}

To assess the impact of a lower-mass binary, we computed a new grid of models using the parameter set AV10 but with $M_1 =1.2$\msun{} and $M_2 =0.8$\msun. Because of its lower mass, the primary reaches a smaller radius at the AGB tip ($\approx 250$\rsun{} compared to 350\rsun{} for $M_1=2$\msun), implying that RLOF could occur at periods between 2000~d and 3000~d, depending on eccentricity. To consider these cases, we extended the range of initial periods by computing models with $P_i =$ 2000~d and 3000~d.

In the \elogP{} diagram, the tracks (Fig.~\ref{fig:e_logP_CRAP}) follow an evolution similar to the 2.0+1.0\msun\ binaries (top panel) and can still be explained in terms of the detailed analysis performed in Sect.~-\ref{sec:detailed_system}.
However, because the evolution of the lower-mass primary is slower and the mass loss rate is smaller, the orbital shrinkage is happening more gradually, leaving more time for the CB disk to act before the tides take over. This translates into a faster increase in the eccentricity with decreasing period (Fig.~\ref{fig:e_logP_CRAP}). This also explains why most binaries that fill a significant fraction of their Roche lobe (i.e. $P_i \leq 4000$~d, orange squares) circularize before the start of RLOF.

Binaries with $P_i \geq 6000$~d do not initiate RLOF and remain on eccentric orbits (red circles). The eccentricity increases due to the CB disk and climbs up to 0.1-0.4, before $\dot{e}_\CB$ saturates. Because of the lower amount of mass available to feed the CB disk,
the surface density $\Sigma_\inn$ reaches approximately half the value ($1-5$~\gcm) of the $M_1=2.0$\msun{} model.

\subsection{Orbital evolution during the post-AGB phase}
\label{sec:evolution_Post-AGB}

Our simulations stop at the end of the AGB phase when less than $\approx 0.05$\msun{} is left in the envelope. At that time, mass transfer by RLOF has ceased and tidal interactions have become very weak. However, the surface disk density $\Sigma_\inn$ is still high and until the disk dissipates [after $\sim 120,000$ years for $P_\orb=2000$~d (Eq.~\ref{tau_visc})], resonant interactions can still operate.
To assess the impact of the CB disk during the post-AGB evolution, we resumed the simulations from the last computed model, freezing the structure (and mass) of the primary, letting the companion star evolve  normally. The simulations are run for an additional $10^6$~yr until the disk has dissipated and the star stabilized in the \elogP{} plane. We consider two cases characterized by different CB disk efficiencies ($\xi$) and in a  first step, neglect the effects of tides on the orbital evolution.

\begin{figure}
\includegraphics[width=\columnwidth]{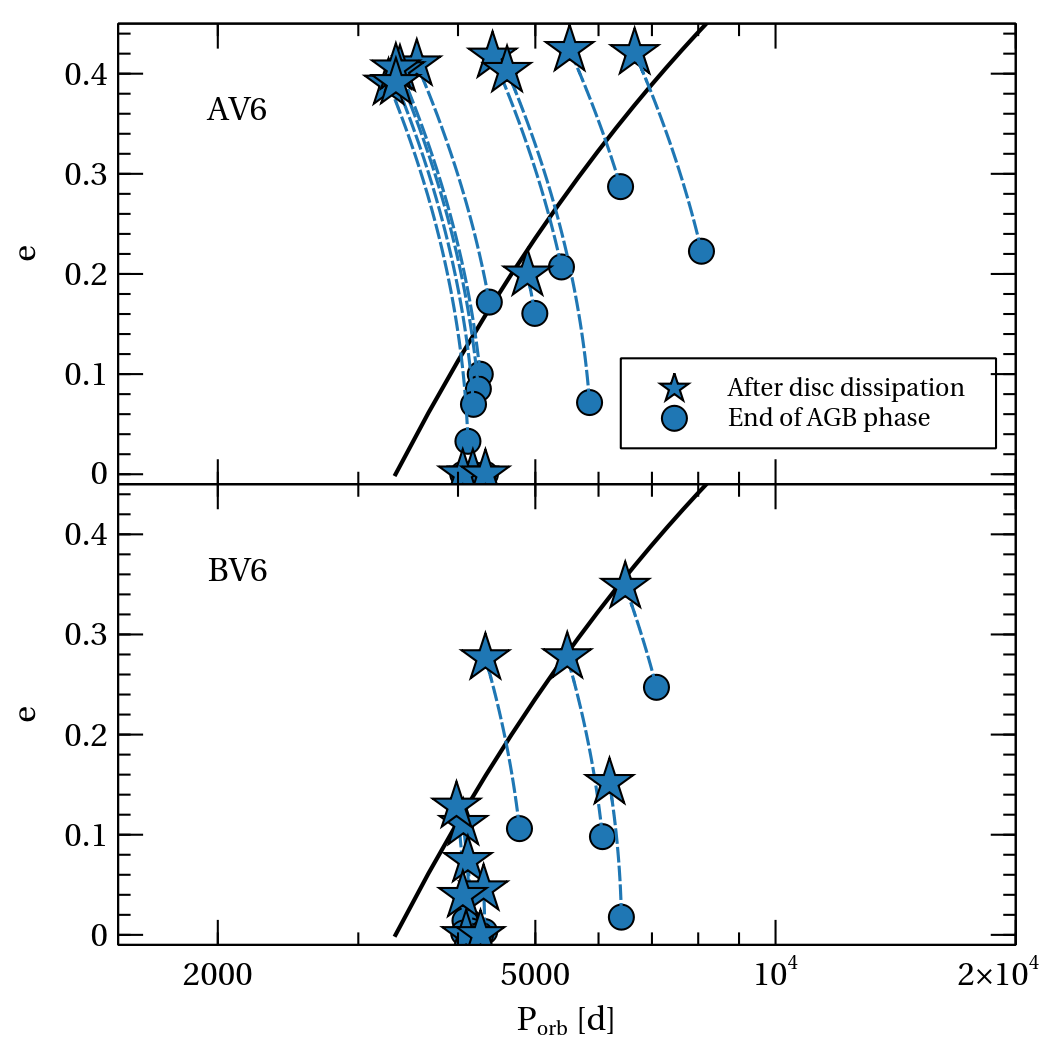}
\caption{Orbital parameters of the 2.0+1.2\msun{} binaries at the end of the simulations, at the AGB tip (circles), for the parameter sets AV6 (top) and BV6 (bottom), and after the dissipation of the CB disk (blue stars).}
\label{fig:elogP_END}
\end{figure}

The  orbital parameters of systems that ended their AGB phase on circular orbits  remain almost unaffected during the disk dissipation phase (i.e., superimposed blue dots and stars with $e\approx 0$ in Fig.~\ref{fig:elogP_END}). For the AV6 systems finishing their AGB evolution in eccentric orbits (blue dots, top panel), an efficient $e$-pumping operates during the post-AGB phase leading to eccentricities $e \approx 0.4$ at which point the resonant interactions saturate (Eq.~\ref{e_damp}). The orbital periods slightly decrease but remain too long to account for the distribution of eccentric barium stars with $P_\orb < 3000$~d.
In models BV6 (having a reduced $\xi$), the CB disk has a weaker effect and the final eccentricities distribute more evenly between 0 and 0.35 with little change in the period (bottom panel).

When the simulations are resumed, the AGB star has a radius of approximately 200\rsun, suggesting that tides, at least at the beginning of the post-AGB phase, could still be effective. While our approach does not account for the peel-off and the reduction of the primary radius during the post-AGB phase, we can estimate the maximum effect of tides by re-running the simulations keeping the tidal torques of the primary to its value at the start of the post-AGB phase. These calculations show almost no changes in the final orbital parameters compared to models without tidal torques. So we conclude that during this fast-evolving phase, within our formalism, tides are negligible whereas the CB disk can have a strong impact.

\section{Discussion}

\subsection{Model limitations}
\subsubsection{WRLOF}

Our paradigm rests on the assumption that WRLOF leads to additional extraction of angular momentum from the system and to the formation of a CB disk. The activation of WRLOF depends on the location of the dust-forming radius with respect to the Roche radius but also on the type of dust to be considered. In our simulations, the primary star remains oxygen-rich. This is a consequence of our use of a relatively low-mass primary, of the absence of extra-mixing processes (for example, overshooting at the base of the envelope), and/or because of a potential increase in the mass loss rate due to tidal interactions or RLOF. If the star turns C-rich, the change in condensation temperature will produce a drop in the dust radius by a factor $\sim0.36$ (Eq.~\ref{eq:Rdust}), potentially bringing $R_\dust$ within the AGB-star Roche lobe. With the deactivation of WRLOF, the Jeans mass loss would be restored and the orbit would widen upon mass loss. The C enrichment of the envelope by third dredge-up episodes will thus hinder the formation of short-period systems. However, the effect may be limited considering that not all stars become C-rich, and when it happens, it can be late in the evolution, with little time left for orbital changes. \cite{Marini2021} find that a 2\msun, $Z=0.008$ star spends 25\% of its AGB lifetime as C-rich, this percentage dropping to 8\% for a 1.25\msun\ star.

The efficiency of WRLOF also depends on the specific angular momentum carried away by the escaping material, as described by our parameter $\eta$, which remains uncertain. In our study, we use the prescription of  \citet{saladino2019}, derived from SPH simulations of circular systems. 
As shown later by \cite{saladino2019ecc}, introducing eccentricity in the simulation reduces $\eta$ by up to 20\%  compared to the circular case. These simulations however do not include dust which presence can have a strong impact on the flow, as it can provide additional acceleration, via radiation pressure on the grains. Dust may as well induce heat transfer as gas and dust may not be at the same temperature, which can potentially impact dust formation. Other physical ingredients, such as the equation of state  \citep[e.g.,][]{TheunsI,Malfait2024} or wind acceleration mechanisms \citep[e.g.,][]{Esseldeurs2023} may potentially impact the parameter $\eta$ as well.

\subsubsection{Disk formation and evolution}

This theory of  excitation of resonances by the CB disk was originally developed within the context of satellites embedded in a circumplanetary disk.  It assumes extreme mass ratios ($q=M_{\mathrm{S}}/M_{\mathrm{P}}\ll 1$, where $M_{\mathrm{P}}$ and $M_{\mathrm{S}}$ are the masses of the planet and satellite, respectively), along with the fact that particles within the disk follow circular near-Keplerian orbits, that the disk is axisymmetric and the eccentricity $e\ll 1$ \citep[see][for a review]{1982ARA&A..20..249G}.

SPH simulations performed by \citet{2012ApJ...749..118S} of a CB disk around an equal-mass binary showed that these assumptions break down. Indeed, they find that the CB disk becomes eccentric, in turn generating an asymmetric feature consisting of a high-density ``lump''. Additionally, from the simulations they calculated the total torque acting on the disk via the binary, and compared this to the total torque
associated with the OLR at the disk's inner edge, as predicted by the linear perturbation theory (i.e. our
$\dot{J}_{m\ell}^{+\mathrm{LR}}$). They found that the latter is about a factor of 4 larger than the simulated values. Since our progenitor binaries have $q\approx 1$, we may therefore overestimate the rate of eccentricity pumping by the same factor (since $\dot{e}_{m\ell}\propto \dot{J}_{m\ell}$). More theoretical work is therefore needed to extend the linear perturbation framework to the binary regime.

Simulations  \citep{2012A&A...545A.127R, miranda2017} also show the appearance of accretion streams emanating from the disk's inner edge, which penetrate the binary system and deposit mass and angular momentum onto the stars \citep[see also][]{1994ApJ...421..651A}. \citet{2012ApJ...749..118S} found that mass transfer could compensate for the orbital shrinkage produced by the binary's interaction with the CB disk, although the orbit contracted overall. Similar results were found by \citet{2012A&A...545A.127R} and \citet{Valli2024}. Whether the orbit expands or contracts, however, depends on the competition between the second, third and fourth terms in Eq.~\ref{eq:adot_a}. In neither of these studies did the accretion streams alter the eccentricity, as also verified by \citet{2012ApJ...749..118S} using analytical arguments.

It is hard to quantify how much mass is exchanged between the binary and the CB disk and how it impacts our results. For post-AGB systems, with CB disks of masses $\lesssim 10^{-2}$\msun, \cite{oomen2020} found that the accretion rates from the CB disk to the binary that reproduce observed depletion patterns, are not strong enough to significantly alter the orbit.
Although it is beyond the scope of this study to investigate the impact of these streams, it is clear that they can substantially modify the evolution of the orbital separation.

\subsubsection{Roche lobe overflow}
\label{discussion:RLOF}

The occurrence of RLOF mass transfer in eccentric orbits can have significant consequences, because the assumptions of the Roche models break down. \citet{2007ApJ...660.1624S} showed that the Roche radius is modified if the star is not synchronized and/or the orbit is eccentric. If the star is rotating sub-synchronously  ($f_{\mathrm{rot}}\equiv \Omega_{*}/\omega_{\mathrm{peri}}<1$, where $\Omega_{*}$ is the star's spin angular speed and $\omega_{\mathrm{peri}}$ is the orbital angular speed at periastron), the Roche radius $R_{\Lone}$ is larger compared to the classical value given by \citet{1983ApJ...268..368E}. The opposite is true for a super-synchronously rotating star ($f_{\mathrm{rot}}>1$). The deformation of the equipotentials opens the Roche lobe around the companion, potentially leading to nonconservative mass transfer through $\Ltwo$. SPH simulations by \cite{lajoie2011} estimated a constant escape rate of 5\% from $\Ltwo$. The escaping matter carries angular momentum, which leads to changes in the orbital response of the binary to the RLOF, and possibly affects the stability of mass transfer.

With our $\eta$ prescription for the angular-momentum loss from the system, the binary systems can undergo RLOF with initial periods $\lesssim 8000$~d (for $v_\wind \approx 10$~\kms). In their study of blue lurkers and blue stragglers, \cite{sun2024} implemented wind accretion by WRLOF following the prescription of \cite{Abate2013}, but assumed that the matter escapes in the Jeans mode, $\eta = \eta_\iso$. As a result of this different treatment of angular momentum loss, they find that RLOF can only be triggered in 2.0+1.1\msun{} systems with initial periods shorter than about 900~d. The condition for RLOF thus depends critically on the treatment of $\eta$.

The RLOF episodes are stable in our simulations. Usually, it is assumed that upon mass loss, convective AGB envelopes expand adiabatically faster than the Roche radius, leading to unstable mass transfer if $q>0.8$ \citep{ivanova2013}. Recent studies \citep[e.g.][]{temmink2023} showed that the adiabatic response of the donor occurs only above a critical mass-transfer rate. Below this threshold, the superadiabatic surface layers can adjust thermally, resulting in self-regulating mass transfer. Performing the same analysis as in \cite{temmink2023} for our 2\msun{} star at the tip of the AGB, we find that this critical rate is as high as $10^{-1}\Myr$, similar to what was found by \cite{ivanova2013} and much larger than the mass transfer rate in our simulations. We also checked that the AGB star does not overflow through the outer Lagrangian point and that upon accretion, the companion does not fill its own Roche lobe, thus avoiding evolution to a contact system.

\subsection{Comparison to observations}

\begin{figure*}
\begin{center}
\includegraphics[width=\textwidth]{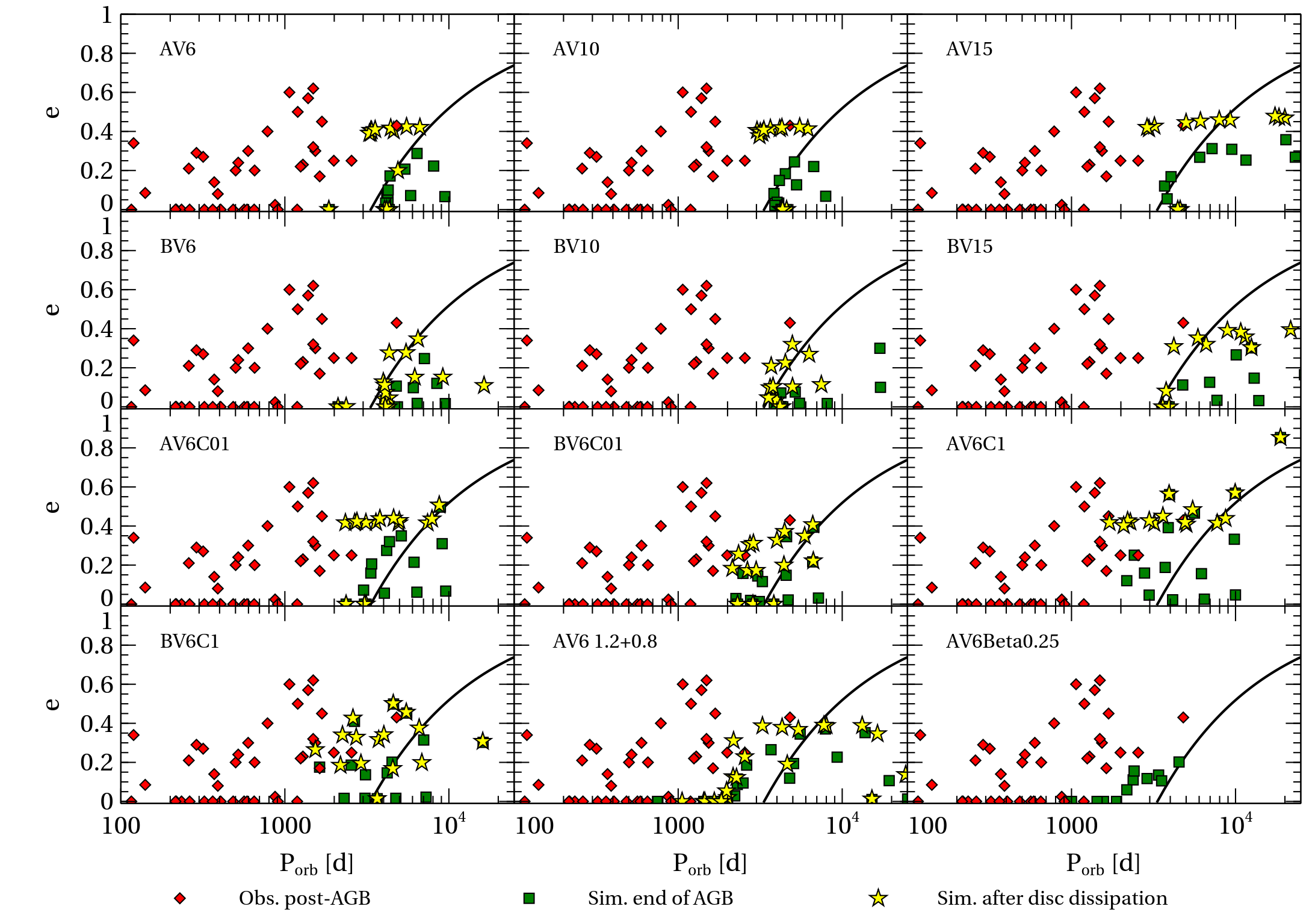}
\caption{Comparison of the final orbital parameters obtained in our simulations with observed post-AGB orbital elements (red diamonds; see Sect.~\ref{sec:sample} for references of the data). Each panel corresponds to a different model.
Green squares correspond to the orbits of the simulated systems at the end of the AGB phase. During the subsequent post-AGB phase, the CB disk continues to act until its dissipation, producing final orbits represented by yellow star symbols.}
\label{fig:observations_recap}
\end{center}
\end{figure*}

Figure~\ref{fig:observations_recap} compares the final orbital parameters of our simulated binaries with the observed values for post-AGB stars. 
There is an almost total incompatibility between the regions of the \elogP{} diagram reached by our simulations and that populated by post-AGB systems. The situation would be somewhat more favorable if the simulations were compared to the barium dwarf sample (see Fig.~\ref{fig:obs}), because the post-AGB sample lacks long-period systems (be it a real lack or the result of insufficiently long radial-velocity monitoring). However, this comparison was not displayed on Fig.~\ref{fig:observations_recap} because the observed properties of the two samples look different in Fig.~\ref{fig:obs}, and it is currently unknown whether this difference is the result of an observational selection bias, of the signature of a further (unidentified) physical process operating between the post-AGB and barium-dwarf phases, or a genuine difference hinting at different progenitor populations. Indeed, binary post-AGB stars are often not s-process rich \citep{2003ARA&A..41..391V}.
In any case, the observed systems with $P_\orb < 2000$~d cannot be explained by our models, because after mass reversal, RLOF leads to an increase in the orbital period.

Our models with tidally enhanced winds (labeled using C01 or C1 suffix in Fig.~\ref{fig:observations_recap}; also Fig.~\ref{fig:e_logP_CRAP}) produce the shortest periods because they do not expand as much. Non-conservative evolution (Fig.~\ref{fig:elogP_NC}) is able as well to produce tight orbits if a lot of mass is ejected during RLOF ($\beta_\RLOF \lesssim 0.5$), but it is still insufficient to account for the observed distributions. In conclusion, another mechanism must be invoked to account for systems with $P_\orb < 2000$~d and the best candidate remains common envelope or grazing-envelope evolution \citep{Kashi2018}. As revealed by Fig.~\ref{fig:observations_recap}, the sought mechanism should be able to produce systems with periods in the range 100 - 2000~d. 
A CB disk, either already present or formed as a result of mass loss during the CE evolution, could subsequently continue to drive eccentricity growth. However, if a system is circularized at the end of the CE phase, reproducing the observed eccentricities of post-AGB binaries would require CB disk masses exceeding those inferred from observations, as noted by \cite{Rafikov2016} and \cite{oomen2020}. This suggests that at least a fraction of post-CE systems must retain some degree of eccentricity.
\begin{figure}[t]
\includegraphics[width=\columnwidth]{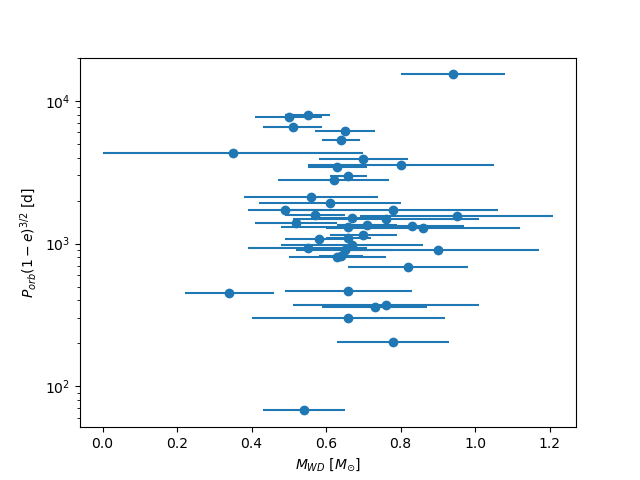}
\caption{Orbital period-white dwarf mass relation from a sample of 60 Ba stars analyzed by \cite{Escorza2023}. The factor $(1-e)^{3/2}$ has little impact and is introduced to determine the orbital period if the system had a separation given by the periastron distance where RLOF is most likely to occur. No clear correlation is present unlike in sdB and He white dwarf binary systems.}
\label{fig:Mwd_Porb}
\end{figure}

Observations have revealed the existence of a correlation between the mass of the helium white dwarf and the orbital period for helium white dwarfs and wide sdB binaries formed during the red giant branch phase which has been interpreted as a result of the strong dependence of a red giant's radius on its core mass \citep{Rappaport1995}. 
To investigate if such a relation exists among Ba stars, as some of these systems may have formed by RLOF, we used the white dwarf masses of barium star binaries with [Fe/H] > -0.5 determined by \cite{Escorza2023}. As shown in Fig.~\ref{fig:Mwd_Porb}, there is no clear evidence of a relation between the WD mass and the orbital period. Two factors could explain this observation : (1) the occurrence of thermal pulses during the AGB phase produces a strong increase in the stellar radius (Fig. 2), triggering episodic RLOF potentially followed by a CE evolution that can  impact the stellar masses and the orbital period and (2) the CO core mass is not a simple function of the stellar radius as it depends on the efficiency of the third dredge-up \citep{Herwig1998}, which in turn depends on internal mixing processes (potentially altered by the presence of the companion) and on the mass loss rate which is impacted by the envelope enrichment following the third dredge-up episodes. The situation is clearly more complicated when dealing with an AGB star and this may reflect in the (absence of a) white dwarf mass–orbital period relation.

\section{Summary and conclusion}
In this study, we explored the effects of WRLOF, CB disk, and tidally enhanced winds primarily on 2.0+1.0\msun\ binaries, with initial periods between 4000 and 16000~d.
We found that WRLOF is initiated by the binary systems with initial periods up to 20000~d as the primary ascends the AGB, however for $16000$~d $<P_i<20000$~d, WRLOF occurs only during thermal pulses, which does not alter the orbital evolution. In this regime, winds escaping the systems efficiently extract angular momentum from the binary system and are the main driver for orbital contraction. This effect depends on the AGB wind speed, with lower values leading to faster orbit contraction. For typical AGB velocities (6-15~\kms), stars initiate RLOF for periods as long as 8000~d. WRLOF may also be initiated during the RGB phase of the primary. However, because of the lower mass-loss rate, its effect on the orbit is limited. By the end of the simulations,  the companion has accreted between 0.2 and 0.5\msun{} by wind and RLOF.

In our model, the CB disk is fed when WRLOF starts. Its effects are strongest during the TP-AGB phase, where the binary gains most of its eccentricity from the CB disk. During the superwind phase, a higher mass loss rate leaves less time for the CB disk to act.

Binaries may initiate RLOF when they are in an eccentric orbit. The phase-dependent mass transfer produces a significant pumping of eccentricity up to 0.4, even if the orbit is nearly circularized. RLOF is stable for most of our binaries because the mass transfer remains below the threshold value ($\sim10^{-1}\;\Myr$) for an effective adiabatic response of the AGB envelope. The few exceptions are for shorter periods and higher eccentricities (typically for the 2.0+1.0\msun{}, $P_i = 4000$~d, $e=0.3$ models). On the other hand, RLOF can also lead to an increase in the orbital period once the mass ratio is inverted. Therefore, systems that initiate (stable) RLOF cluster at the critical period of $\sim 4000$~d, for a wide range of eccentricities.

We explored the effect of tidally enhanced winds on the orbit. The increased mass-transfer rate accelerates the evolution of the AGB star, leaving less time for tidal forces to circularize the orbit. For $B_\wind \gtrsim 10^{-3}$, RLOF occurs in an eccentric orbit and is further pumped during mass transfer with eccentricities reaching $e \approx 0.5$. The accelerated evolution results in fewer thermal pulses which in turn limits the pollution of the AGB envelope by heavy elements and potentially causes difficulties in accounting for their large overabundances observed in the companion barium star. We also simulated several systems where tidal forces were weakened by a factor of 10. The majority of these systems did not circularize before RLOF, leading to similarly high final eccentricities. Introducing non-conservative RLOF allowed to reduce the period while maintaining some non-null eccentricity but this requires a low accretion efficiency which may not be realistic.

Finally, we also studied the effect of the disk during the post-AGB phase by freezing the structure of the primary and following the subsequent evolution of the system. We find that the action of the CB disk on the eccentricity is decisive during the post-AGB phase and partially shadows the previous evolution.

Our models successfully reproduce the observed orbital parameters of dwarf barium stars in
the \elogP{} diagram only for systems with $P_\orb \gtrsim 4000$~d. Specifically, models
incorporating tidally enhanced winds or non-conservative RLOF generate eccentric barium stars
with periods down to 2000~d, somewhat extending the range of accountable eccentric systems.
However, our current models cannot reproduce the eccentric dwarf Ba stars with $P_\orb \lesssim 2000$~d, suggesting that these systems underwent unstable mass transfer followed by common-envelope (CE) evolution. In a recent study, \cite{2021MNRAS.507.2659G} showed that common envelope evolution of initially eccentric binaries can indeed produce post-CE binaries with relatively high eccentricities, up to $e=0.4$. We could also speculate that some fraction of the ejected CE could form a circumbinary disk, that could then further pump the eccentricity. This remains however to be explored with dedicated simulations.

\begin{acknowledgements}
LS is senior FNRS research associate.
\end{acknowledgements}

\bibliographystyle{aa}
\bibliography{references}

\appendix

\section{Variation of $\dot{e}$ and $\dot{a}$ via a circumbinary disk}
\label{app_A}
In this section, we give a more general overview of the resonant interactions than is found in \cite{1979ApJ...233..857G} and \cite{1994ApJ...421..651A}.
We present the equations for the torque between the binary and the CB disk and the ensuing variations of $\dot{e}$ and $\dot{a}$ due to Lindblad resonances in the form usually found in the literature. The evaluation of these equations for the resonance $(2,1)$ gives Eqs.~\ref{Jdot_ml}, \ref{adot_a_CB} and \ref{edot_2}, which correspond to $\dot{J}_\res$, $\dot{a}_\res$ and $\dot{e}_\res$.

In the orbital plane of the binary, centered at the mass center, we expand the time varying potential of the binary $\Phi(r,\theta,t)$ in double series as
\begin{equation}
  \Phi(r,\theta,t)=\sum_{m\ell}\phi_{m\ell}(r)\exp[i(m\theta-\ell\Omega_{\mathrm{B}}t)],
\label{phi_appendix}
\end{equation}
where  $(r,\theta)$ are the polar coordinates, $l$ is the time-harmonic number, $m\geq 0$ is the azimuthal number, $\phi_{m\ell}(r)$ is the radially varying potential component for a given $(m,\ell)$ and $\Omega_{B} = 2\pi/P_{\orb}$ is the binary's mean orbital angular speed, with $P_{\orb}$ the orbital period.
Each component $\phi_{m\ell}(r)$ perturbs the disk's structure, giving rise to an $m$-armed spiral that rotates with a pattern speed
\begin{equation}
  \Omega_{\mathrm{P}}=\left(\frac{\ell}{m}\right)\Omega_{\mathrm{B}},
\label{Omega_P}
\end{equation}
i.e. the spiral is stationary with respect to a frame that rotates with angular speed $\Omega_{P}$. Lindblad resonances (LRs) are localized where the particle of orbital velocity $\Omega(r)$ completes one epicyclic orbit of frequency $\kappa(r)$ between each encounter of the $m$-armed spiral,
\begin{equation}
m[\Omega(r)-\Omega_{\mathrm{P}}]=\pm \kappa(r),
\label{lindblad}
\end{equation}
where henceforth, the upper (plus) sign corresponds to the outer LR (OLR), and the lower (minus) sign to the inner LR (ILR).
Assuming that the particle is following a near-Keplerian orbit ($\kappa(r) \approx \Omega(r)$), then From Eq.~(\ref{lindblad}), we find that the OLR and ILR are located at
\begin{equation}
  r_{\pm\mathrm{LR}}=\left(\frac{m\pm1}{\ell}\right)^{\frac{2}{3}}a.
\end{equation}
For the $(2,1)$ OLR, the resonance is at a radial distance $r_\res = 3^{2/3} a \approx 2.08 a$.

The angular momentum flux at the LRs, $\dot{J}_{m\ell}^{\pm\mathrm{LR}}$ was determined by \citet{1978Icar...34..240G,1979ApJ...233..857G} using linear perturbation theory. The model assumes that (i) the disk is thin ($0.01\lesssim$ $H/r \lesssim 0.1$, where $H$ is the disk's pressure scale height), (ii) the disk is co-planar with the binary orbit and (iii) the potential of the non-axisymmetric perturbations of the disk's structure is small compared to that of the binary.
They find that
\begin{equation}
  \dot{J}_{m\ell}^{\pm\mathrm{LR}}=-m\pi^{2}\Sigma(r_{\pm\mathrm{LR}})|\Psi|^{2}\left(r\frac{\mathrm{d}\mathcal{D}}{\mathrm{d}r}\right)^{-1},
\label{Jdot_ml_general}
\end{equation}
where $\Sigma(r_{\pm\mathrm{LR}})$ is the disk's surface density at the LR, $\mathcal{D}=\kappa^{2}(r)-m^{2}[\Omega(r)-\Omega_{\mathrm{P}}]^{2}$ measures the ``distance''  from the LR, and $\Psi=(\Lambda_{m\ell}\mp 2m)\phi_{m\ell}$ is the forcing function, with $\Lambda_{m\ell}\equiv(\mathrm{d}\ln\phi_{\mathrm{ml}}/\mathrm{d}\ln r)_{\mathrm{LR}}$, also evaluated at the LR \citep{1994ApJ...421..651A}.
For a Keplerian disk, we have
\begin{equation}
r\frac{\mathrm{d}\mathcal{D}}{\mathrm{d}r}=\mp 3\Omega_{\mathrm{B}}^{2}\left(\frac{\ell^{2}}{m\pm 1}\right).
\label{rdD_dr}
\end{equation}

In order to compute the torque for the $(2,1)$ resonance, \cite{1994ApJ...421..651A} use a Taylor expansion for small eccentricities of the potential components
\begin{equation}
\phi_{21}(r)=\frac{9GMa^{2}}{4r^{3}}e\mathscr{M}(1-\mathscr{M}),
\label{Phi_21}
\end{equation}
where $\mathscr{M} = M_1/M$.
The torque $\dot{J}_\res$ (Eq.~\ref{Jdot_ml}) is obtained by evaluating Eq.~\ref{Jdot_ml_general} for $(m,l)=(2,1)$.

We derive now the rates of change of $a$ and $e$ due to the resonant
interaction with a circumbinary disk. The total orbital energy of the
binary system, $\mathcal{E}$, is
\begin{equation}
  \mathcal{E}=-\frac{GM\mu}{2a},
\label{E}
\end{equation}
where $M=M_{1}+M_{2}$ is the total mass and $\mu=M_{1}M_{2}/M$ is the
reduced mass. In case of a resonant interaction and in the absence of
  mass transfer, the rates of change of the semi-major axis,
$\dot{a}$, and of the orbital energy $\dot{\mathcal{E}}$ are related by
\begin{equation}
  \frac{\dot{a}}{a}=-\frac{\dot{\mathcal{E}}}{\mathcal{E}}.
\label{adot}
\end{equation}
We can express $\dot{\mathcal{E}}$ in terms of the rate of change of the
orbital angular momentum, $\dot{J}_\orb$ by using the Jacobi
constant $\mathscr{C}=\mathcal{E}-\Omega_{\mathrm{P}}J_\orb$
where $\Omega_{\mathrm{P}}=(\ell/m) \: \Omega_{\mathrm{B}}$ is the pattern
speed. Hence,
\begin{equation}
   \dot{\mathcal{E}}=\Omega_{\mathrm{P}}\dot{J}_\orb=-\Omega_{\mathrm{P}}\dot{J}_{\mathrm{m\ell}}^{\pm\mathrm{LR}},
\label{Edot}
\end{equation}
where we have made use of the fact that
$\dot{J}_\orb=-\dot{J}_{m\ell}^{\pm\mathrm{LR}}$. Therefore, Eq.~\ref{adot}
writes as
\begin{equation}
  \frac{\dot{a}}{a}=\Omega_{\mathrm{P}}\frac{\dot{J}_{m\ell}^{\pm\mathrm{LR}}}{E_{\mathrm{B}}}.
\label{adot2}
\end{equation}
Substituting Eqs. (\ref{Jdot_ml_general}), (\ref{rdD_dr}) and (\ref{E}) into
Eq.~\ref{adot2} yields
\begin{equation}
 \left(\frac{\dot{a}}{a}\right)_{m\ell}=\mp\frac{2\pi^2}{3}\frac{\Sigma(r_{\pm\mathrm{LR}})}{\mu}\left(\frac{m\pm 1}{\ell}\right)\frac{|(\Lambda_{m\ell}\mp 2m)\phi_{m\ell}|^{2}}{\Omega_{\mathrm{B}}^{3}a^{2}},
\label{adot_a_CB_general}
\end{equation}
which, evaluated at the $(m,l)=(2,1)$ resonance, gives Eq.~\ref{adot_a_CB}.
The eccentricity can be expressed in terms of $J_\orb$ and
$\mathcal{E}$ using
\begin{equation}
  e^{2}-1=\frac{2\mathcal{E}J^{2}_\orb}{(GM)^{2}\mu^{3}},
\label{e2}
\end{equation}
Taking the time derivative of Eq.~\ref{e2} yields
\begin{equation}
  \frac{e\dot{e}}{1-e^{2}}=-\frac{1}{2}\frac{\dot{\mathcal{E}}}{\mathcal{E}}-\frac{\dot{J}_\orb}{J_\orb}=-\frac{1}{2}\frac{\dot{\mathcal{E}}}{\mathcal{E}}+\frac{\dot{J}_{m\ell}}{J_\orb},
\label{ecc_dot}
\end{equation}
where $J_\orb$ is
\begin{equation}
J_\orb=\mu[GMa(1-e^{2})]^{\frac{1}{2}}.
\label{Jorb}
\end{equation}
Inserting Eqs. (\ref{E}), (\ref{Edot}) and (\ref{Jorb}) into
Eq.~\ref{ecc_dot} gives
\begin{equation}
  \frac{e\dot{e}}{1-e^{2}}=-\frac{\dot{J}_{m\ell}^{\pm\mathrm{LR}}}{\mu\Omega_{\mathrm{B}}^{2}a}\left[\frac{\ell}{m}-\frac{1}{(1-e^{2})^{\frac{1}{2}}}\right].
\label{ecc_dot2}
\end{equation}
Finally, inserting Eq.~\ref{Jdot_ml_general} into Eq.~\ref{ecc_dot2} and
using Eq.~\ref{adot_a_CB_general} yields
\begin{equation}
\dot{e}_{m\ell}=\pm\left(\frac{1-e^{2}}{e}\right)\left(\frac{m}{2\ell}\right)
\left[\frac{l}{m}-\frac{1}{(1-e^{2})^{\frac{1}{2}}}\right]\left(\frac{\dot{a}}{a}\right)_{m\ell}.
\label{edot_2_general}
\end{equation}
Equation (\ref{edot_2}) is then obtained by evaluating for the $(2,1)$ resonance.
\clearpage
\end{document}